\documentclass[fleqn,usenatbib]{mnras}

\usepackage{newtxtext,newtxmath}

\usepackage[T1]{fontenc}

\DeclareRobustCommand{\VAN}[3]{#2}
\let\VANthebibliography\thebibliography
\def\thebibliography{\DeclareRobustCommand{\VAN}[3]{##3}\VANthebibliography}

\usepackage{graphicx}	
\usepackage{amsmath}	
\usepackage{color}

\newcommand\ha{H$\alpha$}

\newcommand\nii{[N\thinspace {\sc ii}}

\newcommand\oiii{[O\thinspace {\sc iii}}

\title[The host galaxy of \frb{}]{The most probable host of CHIME FRB 190425A, associated with binary neutron star merger GW190425, and a late-time transient search}

\author[F.~H. Panther et al.]{Fiona H.~Panther$^{1, 2}$, Gemma E. Anderson$^{3}$, Shivani Bhandari$^{4,5,6,7}$, Adelle J. Goodwin$^{3}$, \newauthor Natasha Hurley-Walker$^{3}$, Clancy W. James$^{3}$, Adela Kawka$^{3}$, Shunke Ai$^{8}$, \newauthor Manoj Kovalam$^{1, 2}$, Alexandra Moroianu$^{1, 2}$, Linqing Wen$^{1, 2}$, Bing Zhang$^{8,9}$
\\
$^{1}$Department of Physics, University of Western Australia, Crawley WA 6009, Australia\\
$^{2}$Australian Research Council Centre of Excellence for Gravitational Wave Discovery (OzGrav)\\
$^{3}$International Centre for Radio Astronomy Research, Curtin University, Bentley, WA 6102, Australia\\
$^{4}$ASTRON, Netherlands Institute for Radio Astronomy, Oude Hoogeveensedijk 4, 7991 PD
Dwingeloo, The Netherlands \\
$^{5}$Joint institute for VLBI ERIC, 
Oude Hoogeveensedijk 4, 7991 PD Dwingeloo, The Netherlands \\
$^{6}$Anton Pannekoek Institute for Astronomy, University of Amsterdam, Science Park 904, 1098 XH, Amsterdam, The Netherlands \\
$^{7}$CSIRO Space and Astronomy, Australia Telescope National Facility, PO Box 76, Epping, NSW 1710, Australia \\
$^{8}$Department of Physics and Astronomy, University of Nevada, Las Vegas, NV 89154, USA\\
$^{9}$Nevada Center for Astrophysics, University of Nevada, Las Vegas, NV 89154, USA
}

\date{Accepted XXX. Received YYY; in original form ZZZ}

\pubyear{2022}

\begin{document}

\newcommand{\frb}{FRB~20190425A}
\newcommand{\host}{UGC10667}
\newcommand{\bns}{GW190425}
\newcommand{\figref}[1]{Figure~\ref{#1}}
\newcommand{\tabref}[1]{Table~\ref{#1}}

\label{firstpage}
\pagerange{\pageref{firstpage}--\pageref{lastpage}}
\maketitle

\begin{abstract}
The identification and localization of Fast Radio Bursts to their host galaxies has revealed important details about the progenitors of these mysterious, millisecond-long bursts of coherent radio emission. In this work we study the most probable host galaxy of the apparently non-repeating CHIME/FRB event \frb{} --- a particularly high luminosity, low dispersion measure event that was demonstrated by \cite{Moroianu2022} to be temporally and spatially coincident with the LIGO-Virgo-KAGRA binary neutron star merger \bns{}, suggesting an astrophysical association (p-value 0.0052). In this paper we remain agnostic to this result, and we confirm \host{} as the most probable host galaxy of \frb{}, demonstrating that the host galaxies of low dispersion measure, one-off CHIME FRBs can be plausibly identified. We then perform multi-wavelength observations to characterize the galaxy and search for any afterglow emission associated with the FRB and its putative GW counterpart. We find no radio or optical transient emission in our observations $2.5\,\mathrm{yr}$ post-burst. \host{} is a spiral galaxy at $z\sim0.03$, dominated by an old stellar population. We find no evidence of a large population of young stars, with nebular emission dominated by star formation at a rate of $1-2\,\mathrm{M_\odot\,yr^{-1}}$. While we cannot rule out a young magnetar as the origin of \frb{}, our observations are consistent with an origin in a long delay-time neutron star binary merger as posited by \cite{Moroianu2022}.
\end{abstract}




\section{Introduction}
Fast radio bursts (FRBs) are enigmatic millisecond bursts of coherent radio emission originating at cosmological distances. Since the identification of the Lorimer burst in 2006 \citep{Lorimer2007}, hundreds of FRBs have been identified. Initially discovered serendipitously, dedicated radio surveys such as the Commensal Real-time ASKAP Fast Transients Survey (CRAFT) with the Australian Square Kilometre Array Pathfinder (ASKAP) \citep{Bannisteretal2017}, and the Canadian Hydrogen Intensity Mapping Experiment FRB project (CHIME/FRB) \citep{CHIME2019a} have significantly increased the rate of detection of FRBs. However, despite the apparent ubiquity of these events, what powers these bright flashes of radio emission remains mysterious.

FRBs are broadly classified into two categories: repeating FRBs and apparent non-repeaters. While repeating FRBs must arise from events that do not destroy the central engine, with young magnetars being a lead candidate \citep{michilli18}, non-repeating FRBs may originate from the merger of two compact objects, e.g.\ the inspiral, merger, and/or remnant collapse of a binary neutron stars
\citep[e.g.][amongst others]{Usov2000,Hansen2001,Totani2013,Lyutikov2013,Zhang2014,Wang2016}.
However, it is also possible that apparent non-repeaters are simply too distant or repeat too rarely for more than one burst to be observable \citep{Caleb2019,James2019limits,Gardenier2021}. Whether or not repeating FRBs and apparent non-repeaters are truly different is an open question in the field. Interestingly, bursts from repeating FRBs tend to be broader in time and narrower in frequency (which tends to drift downward with time, known as the 'sad-trombone effect') than apparently once-off bursts, which is suggestive of two source populations \citep{chime_morphology_2021}.

Searching for multi-wavelength and multimessenger signals that originate from the same source as FRBs may reveal their progenitors. \cite{Moroianu2022} posit a common origin of the binary neutron star (BNS) merger \bns, detected by the LIGO-Virgo-KAGRA (LVK) collaboration \citep{abbott2020}, and \frb{}, detected by CHIME/FRB \citep{chime_catalog1_2021}. The time delay between the two events ($2.5\,\mathrm{hrs}$) is consistent with the FRB emission arising from the collapse of a supramassive neutron star or quark star to a black hole, as theorised by \cite{Zhang2014}. \cite{Moroianu2022} claim the sky localization, time offset, and distance to the events is consistent with a common astrophysical origin with a p-value of 0.0052 (2.8\,$\sigma$). Such an FRB would be more likely to escape the BNS ejecta if it is viewed out of the plane of the merger.
 
 A dedicated deep search by the LVK for gravitational wave signals associated with CHIME/FRB events did not find any signal consistent with a common origin with \frb{} in a 12-minute window \citep{2022arXiv220312038T}. An independent search by \cite{OGCFRB} using LVK public data also returned a null result searching for coincidences between CHIME/FRB events and GW triggers identified in the 4th Open Gravitational Wave Catalog \citep{4-OGC} in a $\pm100\mathrm{s}$ coincidence window around the time of each considered CHIME FRB. However, in the case of the progenitor of \frb{} being the collapse of a hypermassive neutron star or quark star which occurs on hour -- rather than minute --  timescales, this does not rule out the chance of astrophysical association between \frb{} and \bns{} proposed in \cite{Moroianu2022}.
 
CHIME/FRB can only localize FRBs with an accuracy of a few arcminutes on average, and the majority of the 535 FRBs observed by CHIME to date have not been localized to individual hosts, with the exceptions being three repeating FRBs from nearby galaxies \citep{Marcote2020,Bhardwaj2021_M81,Bhardwaj2021_NGC3252}. If the association of \host\, with \frb\, is confirmed, this may represent the first identified host galaxy of a one-off CHIME FRB. Moreover, if \bns\, and \frb{} have a common origin, the identified host would be the second galaxy known to have hosted a BNS detected via gravitational waves. 

To date, no transient emission associated with cosmological FRBs has been conclusively identified, however several progenitor models strongly suggest that they may be associated with cataclysmic events such as gamma-ray bursts (GRBs), particularly short-duration GRBs, which are now known to be merging binary neutron stars \citep[for a summary of models see][]{rowlinson19}. These theories have motivated the search for FRB-like signals associated with GRBs as well as GWs  \citep[][]{Bannister2012,Palaniswamy2014,Obenberger2014,Kaplan2015,Andreoni2017,Anderson2018,Callister2019,Rowlinson2019grb,Rowlinson2021,Bouwhuis2020,Anderson2021mwa,Tian2022,Tian2022vcs,Curtin2022} 
so such events must therefore have associated afterglow emission as was the case for GW170817 \citep{GW170817_multi}. One should therefore see the emission associated with an expanding jet if it is optimally oriented, or isotropic emission from an associated kilonova event that would be expected from a binary neutron star merger. FRB emission closer to the merger axis is also less likely to be absorbed by the ejecta material, further motivating the search for a jet. There are a number of strong observational constraints on the association between individual FRBs, GRB prompt emission and afterglows, and SNe. Dedicated multi-wavelength followup of a number of FRBs has revealed these specific events are not associated with GRB- or SNe-like emission \citep[e.g.][]{Petroff2015, Marnoch2020, Nunez2021, gourdji20, Kilpatrick2021, Curtin2022}. Nevertheless, the possibility of binary neutron star mergers as the origin of a sub-population (or subsample) of FRBs has not been conclusively ruled out.

In the absence of the afterglow revealing the central engine that powers the FRB, identifying and characterizing their host galaxies can shed light on their origin. Key insights into other astrophysical transients have been obtained by studying their host environment. For example, transients arising from host galaxies with high star formation rates have been associated with the deaths of young, massive stars \citep[e.g.][]{Taggart2021}. Others may be almost exclusively associated with galaxies that have long since ceased star formation \citep[e.g.][]{panther2019}, indicating they arise from old progenitors with long delay times. Connections between transients and gas-phase metallicity can also yield important information about likely progenitors --- for example, low metallicities hint at the types of massive stars that may be the  progenitors of long-duration GRBs \citep{Perley2016}.

Unlike supernovae, which can be reliably localized within their host galaxies, only a comparatively small number of FRBs have been localized to individual host galaxies \citep{Heintz2020, Bhandari+22}\footnote{An online catalog of known FRB hosts is maintained at http://frbhosts.org}. Very precise localizations, to within milliarcseconds, are typically only known for repeating FRBs \citep[e.g.][]{Niu2021}. This has enabled the detailed study of their local environments within their hosts.

The study of the $\sim 20$ FRB host galaxies that have been identified has demonstrated that they are unlikely to be associated with exclusively old stellar populations ($>5-10\,\mathrm{Gyr}$ old) with no ongoing star formation \citep{Heintz2020, Bhandari+22}. The majority of FRBs -- both repeaters and apparent non-repeaters -- are found in environments with star formation rates of $\sim 0.05 - 10\,\mathrm{M_\odot\,yr^-1}$ \citep{Heintz2020}. Many hosts of apparent non-repeaters reside in the green valley \citep{Bhandari+22}. A significant fraction of apparent non-repeating FRBs are found in galaxies that exhibit low ionization nuclear emission line (LINER) emission, which may indicate that FRBs are more commonly found in galaxies that have recently undergone mergers or contain a large amount of gas that fuels either active star formation or AGN activity. FRB hosts show no trend for low metallicities, suggesting that the progenitors are not tied to extremely metal-poor environments \citep{Bhandari+22}. 

In this paper we present radio and optical observations of the probable host galaxy of \frb{}. We present our identification of the probable host galaxy \host{}, and our observations of \host{} using the Very Large Array (VLA), Murchison Widefield Array (MWA), archival SDSS photometry and spectra, and the ANU 2.3m telescope/Wide Field Spectrograph. We discuss the properties of the host in the context of other known FRB host galaxies, and perform a search for any afterglow emission associated with \frb{} (and the putatively connected \bns). 

\section{Host Galaxy Identification}
\label{sec:hostid}
\cite{Moroianu2022} searched for a host galaxy of \bns\ and \frb\ within the CHIME $1\,\sigma$ uncertainty region of ${\rm RA} = 255.72 \pm 0.14^{\circ}$ and ${\rm DEC} = 21.52 \pm 0.18^{\circ}$. \host{} was identified as the likely host as being the only galaxy within the allowed redshift range of \bns{} from GW data, and the maximum redshift $z_{\rm max} = 0.0394$ allowed from the dispersion measure (DM$=128$\,pc\,cm$^{-3}$) of \frb.

Here, we extend the search for possible host galaxies to include CHIME's full localisation skymap \citep{chime_catalog1_2021}, and apply a Bayesian framework for assigning probabilities to all candidates. Our search is agnostic to any priors set by \bns{}, and considers only the probability that an individual galaxy hosted \frb{}.

As \frb{} is significantly better localized than \bns{}, we focus on identifying the host of \frb{}. This also demonstrates that the host galaxies of low-DM CHIME FRBs can be plausibly identified regardless of whether or not they are associated with another multimessenger signal.

\begin{figure*}
	 		\includegraphics[scale=0.8]{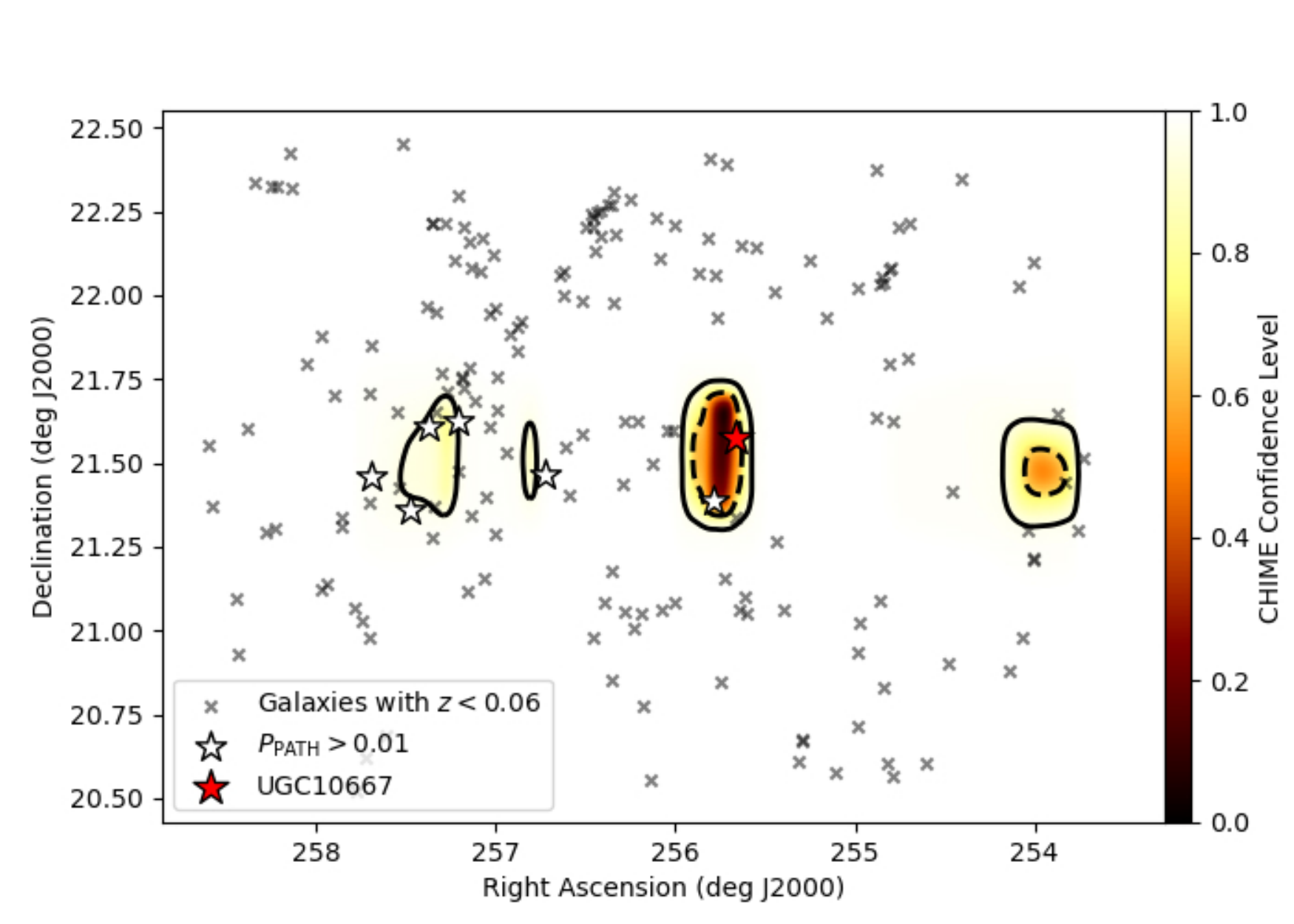}
	 		\caption{\label{fig:Chimemap} All galaxies identified in the host galaxy search (crosses). Galaxies with a PATH association probability $P_\mathrm{PATH}\ge0.01$ are indicated as white stars. \host{}, the most probable host, is indicated with the red star. Color map and contours indicate the CHIME/FRB confidence interval for the location of \frb{}, with dashed contour: CHIME/FRB 68\%, solid: CHIME/FRB 95\%.}
\end{figure*}

\begin{table*}
	\centering
	\caption{Candidate host galaxies of \frb{} with $P_\mathrm{PATH}\ge0.01$. All galaxies lying within the 99\% confidence region of the CHIME localisation, and $z<0.06$, are considered. Given are the object name, RA and DEC, confidence level $\mathrm{CL_{CHIME}}$, r-band magnitude $m_r$, and posterior likelihood $P_\mathrm{PATH}$ as defined by \eqref{eq:posteriors}.
	}
	\label{tab:besthosts}
	\begin{tabular}{lcccccc}
		\hline
		Identifier & RA & DEC & Redshift & $\mathrm{CL_{CHIME}}$	& $m_r$ & $P_\mathrm{PATH}$\\
		\hline
		UGC10667	&               255.66248	& 21.57675 & 0.0312 & 0.513 & 15.031 & 0.787\\
        WISEAJ170310.07+212309.9	& 255.79193	& 21.38602 & 0.0475  & 0.417 & 17.718 & 0.0334\\
        WISEAJ170930.73+213633.8	& 257.37796	& 21.60942 & 0.0487 & 0.936 & 15.959 & 0.0311\\
        WISEAJ170851.16+213718.1	& 257.21318	& 21.62171 & 0.0592 & 0.945 & 15.939 & 0.0276\\
        WISEAJ171046.84+212732.9	& 257.69512	& 21.45914 & 0.0473 & 0.984 & 15.575 & 0.013\\
        WISEAJ170653.30+212744.1	& 256.72211	& 21.46232 & 0.0325 & 0.989 & 15.324 & 0.0122\\
        WISEAJ170954.22+212125.1	& 257.47593	& 21.35695 & 0.0498 & 0.965 & 16.382 & 0.01\\
		\hline
	\end{tabular}
\end{table*}

\subsection{Searching for potential hosts}
\label{sec:hostsearch}
We search for potential host galaxies of \frb{} using the NASA Extragalactic Database (NED)\footnote{The NASA/IPAC Extragalactic Database (NED) is funded by the National Aeronautics and Space Administration and operated by the California Institute of Technology.}. We search for objects classified as extragalactic in the NED catalogs using two main criteria. Firstly, we tile cone searches to cover the entire CHIME localisation region, shown in Figure \ref{fig:Chimemap}. Secondly, we remove galaxies from our sample with $z>0.06$. This corresponds to a maximum luminosity distance of $270.4\,\mathrm{Mpc}$, which is significantly more distant than the maximum distance of 180\,Mpc argued by \cite{Moroianu2022} based on the measured DM of \frb{} of 128\,pc\,cm$^{-3}$.

We only consider objects listed in NED as `galaxies', excluding galactic and extragalactic stars, UV, and x-ray sources with no optical or IR cross-match, as well as objects only identified by the NRAO Very Large Array (VLA) Sky Survey \citep[NVSS;][]{1998AJ....115.1693C},  that have not been cross-matched to a known extragalactic source. Many of these objects lack the information required by the formalism we use to deduce the probability of association.

Catalogs indexed by NED that we utilize for the host galaxy search are:
\begin{itemize}
    \item AllWISE\footnote{https://wise2.ipac.caltech.edu/docs/release/allwise/expsup/}: stellar masses at $0 < z < 0.1$ range from $\log_{10}(M_*) \simeq 7 - 11$. It is noted that the catalog may be incomplete for $\log(M_*)<7$
    
    \item The SDSS DR6 Photometric catalog\footnote{http://cas.sdss.org/dr6/en/}, with an r-band magnitude limit $m_r = 22.2$
    
    \item 2MASS Point Source Catalog\footnote{http://tdc-www.harvard.edu/software/catalogs/tmpsc.html}: NED only lists the 2MASS Point Source Catalog (2MASS PSC), which is predominantly galactic objects. The PSC contains point-source processed photometry for virtually all resolved sources found in the Extended Source Catalog (ESC), as well as more distant, unresolved galaxies. The 2MASS PSC is complete down to $J<15.8$, $H<15.1$ and $Ks<14.3$ mag.
\end{itemize}
While we do not place any magnitude limit on our search using NED, we note that the catalogs indexed by NED for this region of the sky are magnitude-limited, and hence we cannot rule out the origin of \frb{} to be an object below the magnitude limit of these surveys (for example, an ultra-diffuse galaxy).

This search returns 176 unique candidate galaxies.

\subsection{Calculation of host association probabilities}

To evaluate the probability that any given galaxy is the true FRB host, we use the `PATH' methodology of \citet{PATH}. This is a Bayesian formalism which combines the localisation likelihood of an FRB from radio observations with priors on FRB host galaxy magnitudes, radial offsets, and on the chance that the true host is unobserved. In the case of \frb{}, the localisation error is much greater than the extent of the candidate host galaxies. We therefore ignore the potential radial offset of each host. In this simplified formalism, the posterior probability of candidate host $i$, $P_{\mathrm{PATH},i}$, is
\begin{eqnarray}
P_{\mathrm{PATH},i} & = & \frac{p(\Omega_i) p(m_{r,i})}{\sum_j p(\Omega_j) p(m_{r,j})}. \label{eq:posteriors}
\end{eqnarray}
Here, $p(\Omega_i) = 1-\mathrm{CL_{CHIME}}$, where $\mathrm{CL_{CHIME}}$ is the localisation confidence level at the position $\Omega_i$ of the candidate galaxy (see \figref{fig:Chimemap}). The term $p(m_{r,i})$ is a prior on the r-band magnitude of the galaxy, $m_r$, given by
\begin{eqnarray}
p(m_r) & \propto & \frac{1}{\phi(m_r)},
\end{eqnarray}
where $\phi(m_r)$ is the number density of galaxies with magnitude $m_r$ on the sky. We take $\phi(m_r)$ from \citet{driver16}, using the implementation in the \textsc{frb} code \citep{frbgithub}. Effectively, this prior states that the number of FRBs occurring in host galaxies in a given magnitude range is independent of that magnitude, since the product of the number of such galaxies, and the prior on each galaxy, will be unity. Due to the necessary proximity of the host galaxy, we assume that the true host is detectable in optical data, and set the prior on the true host being undetected to be zero. Thus posterior probabilities are normalised in \eqref{eq:posteriors} to sum to unity.

Applying this methodology to the 176 candidate hosts shows that the nominal host galaxy \host{} has a posterior probability of 78.7\% to be the true host galaxy. The second-to-best candidate, WISEAJ170310.07+212309.9 ($P_{\mathrm{PATH}}=3.34\%$), has a significantly higher magnitude and hence lower magnitude prior, while the remaining candidates lie outside the 90\% localisation contours, and have posteriors in the range 1--3\%. All candidates with $m_r > 18$ are found to have negligible posterior probabilities ($P_{\rm PATH} < 4 \cdot 10^{-4}$). Since this is within the limiting magnitude of the surveys used, this justifies our assumption that the true host is visible. If we allow for a small prior probability $p(U)$ on the host being unobserved, the PATH formalism finds a posterior estimate $p(U|{\rm data}) \sim 4.2 p(U)$, with the corresponding posteriors calculated from Eq.~\ref{eq:posteriors} being reduced by a factor of $[1-p(U|{\rm data})]^{-1}$.

Our confidence of 78.7\% of \host{} being the true host is comparable to the most likely host of many other FRBs, although the association is not as certain as that of some \citep{PATH}. We therefore suggest that \host{} is treated as the default host galaxy of \frb.

\section{Observations}
We perform a series of observations to characterize \host{}, including determining the star formation rate, the age of its stellar population and the dominant source of ionizing radiation in the galaxy. We also search for the afterglow associated with \frb{}/\bns{} using our new observations and archival data.  

\begin{table}
	\centering
	\caption{Observations of UGC10667 used in this paper}
	\label{tab:VLAobs}
	\begin{tabular}{c}
	    VLA  \\
	\end{tabular}
	\begin{tabular}{lcccccc}
		\hline
		Date & Array & $\nu$  & $F_{\nu}$ & u($F_{\nu}$)  & Image RMS  \\
		 (UTC) & Config. & (GHz) &  (uJy) & (uJy) & (uJy) \\
		\hline
		\hline
		2017-09-09 & VLASS &3 & $<417$ & - & 139 \\
		2020-08-08 & VLASS & 3& $<441$ & - & 147 \\
		\hline
		2021-09-09 01:51 & VLA-C & 6 & 347 & 57 & 11.2 \\
		\hline
		2021-10-16 20:23 & VLA-B & 1.5 & 1490 & 225 & 27.5 \\
		 & & 3 & 588 & 119 & 12 \\
		 & & 6 & 357 & 29 & 11 \\
		 & & 10 & $<30$ & - & 9.3 \\
		\hline
	\end{tabular}
	
	\begin{tabular}{c}
    MWA
	\end{tabular}
		\begin{tabular}{lccccc}
		\hline
		Date & Array & $\nu$  & $F_{\nu}$ &  Image RMS  \\
		 (UTC) & Config. & (GHz) &  (mJy) &  (mJy) \\
		\hline
		\hline
 		2019-07-02T14:10 & Phase~\textsc{II} extended & 0.199 &  - & 5.3 \\
 		2019-07-03T15:56 & Phase~\textsc{II} extended & 0.154 &  - & 5.5 \\
 		\hline
 		 & combined image & 0.170 & $<12$ & 4 \\
		\hline
	\end{tabular}
	
		\begin{tabular}{c}
    ASKAP
	\end{tabular}
		\begin{tabular}{lcccccc}
		\hline
		Date & Array & $\nu$  & $F_{\nu}$ & u($F_{\nu}$)  & Image RMS  \\
		 (UTC) & Config. & (GHz) &  (uJy) & (uJy) & (uJy) \\
		\hline
		\hline
		2019-04-24 & ASKAP& 0.88 & 2840 & 860 & 260 \\
		\hline
	\end{tabular}
	
		\begin{tabular}{c}
    ANU 2.3m/WiFeS
	\end{tabular}
		\begin{tabular}{lccccc}
		\hline
		Date &  Wavelength  & Gratings & Dichroic  & Exposure\\
		 (UTC) & range (\AA) & - &  - & time (s)  \\
		\hline
		\hline
	    2021-08-31 08:43             
 & 3300 - 9200 & R/B3000 & RT560 & 3 x 1200s \\
		\hline
	\end{tabular}

\end{table}

\subsection{MWA}
We searched the Murchison Widefield Array \citep[MWA;][]{2013PASA...30....7T,2018PASA...35...33W} archive for observations covering \host{} that were taken after the transient event, finding 30$\times$5-minute observations taken at 139--170\,MHz on 2019-07-02 (68 days after \frb) as part of project D0000 (Unspecified Director's Time). We performed direction-independent bandpass calibration using the \textsc{MitchCal} algorithm \citep{2016MNRAS.458.1057O} and a local sky model based on the Galactic and Extragalactic All-sky MWA \citep[GLEAM;][]{2015PASA...32...25W,2017MNRAS.464.1146H} survey, with sources above Declination~$30^\circ$ populated from NVSS \citep{1998AJ....115.1693C} and VLA Low-frequency Sky Survey Redux \citep[VLSSr;][]{2014MNRAS.440..327L}. We visually inspected the gains and applied them. We then used \textsc{WSClean} \citep{2014MNRAS.444..606O} to invert the visibilities and deconvolve each snapshot image, using settings derived from the GLEAM-eXtended (GLEAM-X; Hurley-Walker et al. submitted) survey: a ``robust'' weighting of $+0.5$ \citep{1995AAS...18711202B}, multi-frequency-synthesis imaging across the 30.72\,MHz bandwidth, a pixel scale yielding $>3$~pixels per synthesised beam full-width-half-maximum (FWHM), an image size encapsulating primary beam response $>$10\,\%, cleaning to $3\times$ the local root-mean-square (RMS) noise $\sigma$, and then down to $1\times$ that level at the locations identified as containing components (``auto-masking''). The resulting images were corrected for the primary beam using the Fully-Embedded Element simulations by \cite{2017PASA...34...62S}.

\begin{figure}
	 		\includegraphics[width=\columnwidth]{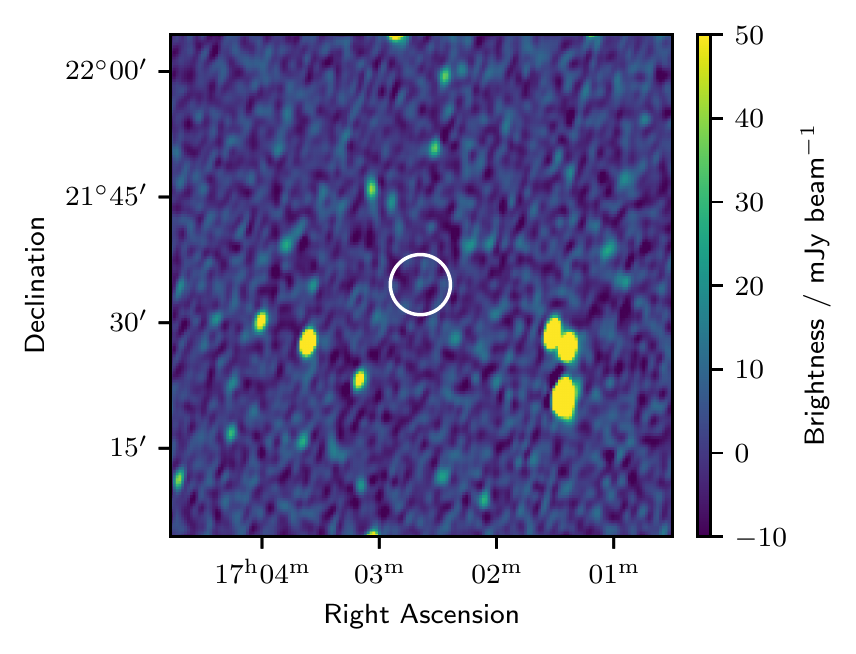}
	 		\caption{\label{fig:MWA}Mosaic of 148~minutes of MWA data recorded on 2019-07-02 at 154\,MHz and 2019-07-03 at 199\,MHz. The RMS noise is 4\,mJy\,beam$^{-1}$. The location of \host{} is indicated with a white open circle.}
\end{figure}

At this stage, we visually inspected each image, finding significant ionospheric distortions to sources in 16~snapshot observations, which were thus discarded. The remaining 14~images were corrected for position shifts using \textsc{fits\_warp} \citep{2018A&C....25...94H}, resulting in an astrometric accuracy of $\sim$5$''$. Using \textsc{flux\_warp} \citep{2020PASA...37...37D}, the mean flux density scale was adjusted to match the initial sky model, a correction of about $20$\,\% for each snapshot, with residual flux density scale uncertainty of $\sim$5\,\%. We combined the corrected snapshots in the image plane using \textsc{swarp} \citep{2002ASPC..281..228B}, resulting in a mosaic with RMS noise of 5\,mJy\,beam$^{-1}$ (Figure~\ref{fig:MWA}). \host{} is not visible in this final image, and we therefore measure an upper 3-$\sigma$ flux density limit of $<$15\,mJy\,beam$^{-1}$ at its location, at 154\,MHz.

For an independent check of our MWA-derived flux density upper limit, we also looked for emission at this location in the Tata Institute of Fundamental Research Giant Metrewave Radio Telescope Sky Survey Alternative Data Release 1 \citep[TGSS-ADR1][]{2017A&A...598A..78I}. This survey is less sensitive to diffuse emission \citep{2017arXiv170306635H} so may not be sensitive to objects such as UGC10667. The local RMS noise is 3\,mJy\,beam$^{-1}$ and no detection was made.

\subsection{ASKAP}
The Rapid ASKAP Continuum Survey \citep[RACS,][]{McConnell2020} also covered the location of \host{} on 2019 April 24 at 880\,MHz, one day before the FRB occurred. In the RACS calibrated data product image, there is a 2.4\,$\sigma$ detection of a point source at the location of \host, with a flux density of 2.84$\pm$0.86\,mJy\,beam$^{-1}$ at 880\,MHz (Table \ref{tab:VLAobs}). Whilst this detection is not statistically significant, it is consistent with the flux density expected at 880\,MHz based on the measured spectral index of the radio emission at 1--10\,GHz (Section \ref{sec:radio}). The galaxy is unlikely to be resolved by RACs at 1.4\,GHz (angular resolution of $\approx$15\,arcsec), and so the apparent 2.4\,$\sigma$ point source detected could indeed be \host.

\subsection{Historical VLA observations}
The NRAO's Karl G. Janksy Very Large Array (VLA) Sky Survey \citep[VLASS,][]{Lacy2020} covered the coordinates of \host{} in both VLASS 1.1 and 2.1 on 2017 September 09 and 2020 August 08 respectively at 3\,GHz. There is no detection of the host in either quick-look image, with a 3\,$\sigma$ upper limit of $<417$\,$\mu$Jy\,beam$^{-1}$ and $<441$\,$\mu$Jy\,beam$^{-1}$ in each observation (Table \ref{tab:VLAobs}). Whilst the host should theoretically have been detected in these observations, we note that the array configuration for VLASS observations is not sensitive to diffuse radio emission such as the resolved emission from \host, instead favouring detection of point sources. Since VLASS has an angular resolution of $\approx$2.5\,arcsec, this would resolve the emission from \host, resulting in a non-detection. 

\begin{figure}
	 		\includegraphics[width=\columnwidth]{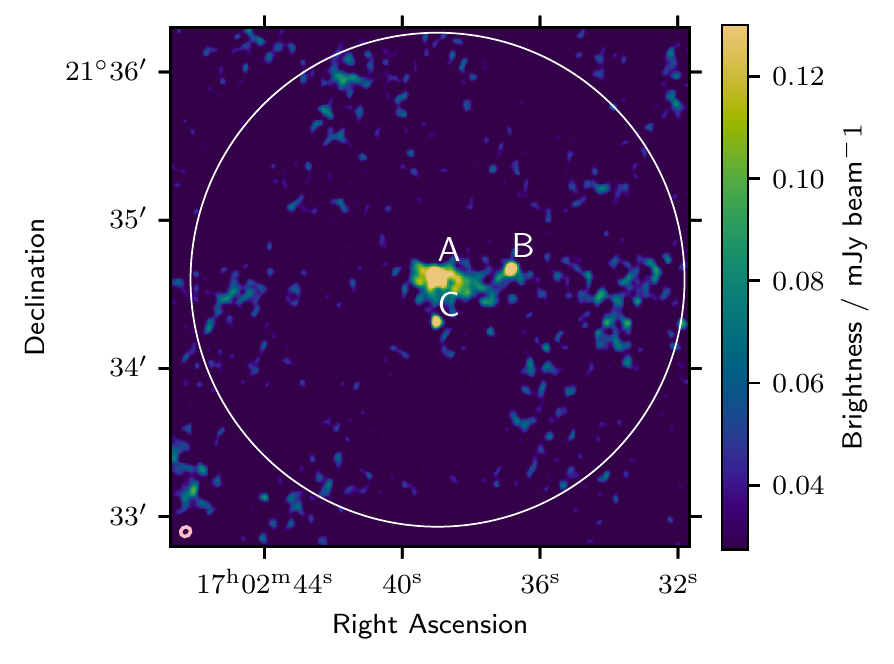} 
	 		\caption{\label{fig:VLA-Lband}VLA 1.5 GHz image of \host{} recorded on 2021-10-16. The RMS noise is 27.5\,$\mu$Jy\,beam$^{-1}$. The beam size is indicated in the bottom corner in pink and the 100\,arcsecond search radius for transient emission is indicated by the large white circle.
	 		The image is labelled as follows; `A': the nucleus of \host; `B': compact flat-spectrum radio source, likely a background AGN associated with SDSS J170237.03+213440.8 and IrS WISEA J170236.90+213440.9; `C': compact steep-spectrum radio source, also likely a background AGN associated with optical source WISEA J170239.00+213419.9 (for details see Section~\ref{sec:radio}).}
\end{figure}

\subsection{Targeted VLA observations}
We obtained radio observations centered on UGC10667 with the VLA on 2021 September 9 and 2021 October 16 (program ID: 21B-343, PI: Goodwin). Since the archival observations showed no radio detection, our initial observation on 2021 September 9 was taken at 4--8\,GHz (C-band) to search for any radio emission from a possible remnant or outflow that may have been produced, as well as any diffuse radio emission from star formation in the galaxy. The galaxy was detected in our observation at 4--8\,GHz, motivating the follow-up observation on 2021 October 16 at 1--12\,GHz (L-, S-, C-, and X-band with central frequencies 1.5\,GHz, 3\,GHz, 6\,GHz, and 10\,GHz) to characterise the radio spectrum of the galaxy and search for any transient emission associated with the event. 

The data were reduced using standard procedures in the Common Astronomy Software Package \citep[CASA 5.6;][]{Mcmullin2007}, including the VLA pipeline. 3C286 was used as the flux density calibrator and J1716+2152 was used as the phase calibrator for all observations and frequencies. Images of the target field were created using the CASA task TCLEAN. The total source flux density was measured in the image plane using the CASA task IMFIT, without fitting a point source due to the extended nature of the emission from the galaxy. The errors reported for the measured flux densities include a statistical uncertainty and a systematic one due to the uncertainty in the flux-density bootstrapping, estimated at 5\%. Where possible, we split each observing band into four sub-bands (two for 10\,GHz) to extract the spectral properties of the sources in the images. The resulting host flux densities and image RMS
are summarised in Table \ref{tab:VLAobs}. A final VLA image of the field created from the 1.5\,GHz observation on 2021-10-16 is show in Figure~\ref{fig:VLA-Lband}, clearly indicating the host galaxy label as `A' and two nearby unresolved sources labelled `B' and `C' (all discussed further in Section~\ref{sec:radio}). Contours of the radio emission
for each frequency band are over-plotted on an archival SDSS image of the field
in Figure \ref{fig:rgbimage}.

\subsection{Optical observations of \host{}}
We obtain integral field unit (IFU) observations of \host{} using the Wide Field Integral Field Spectrograph (WiFeS) instrument \citep{Dopita07} using the Australian National University $2.3\,\mathrm{m}$ Advanced Technology Telescope at Siding Spring Observatory. These observations were obtained on August 31 2021. WiFeS is a wide-field IFU instrument, with a total field of view of $25"\times38"$. The instrument configuration used in this observation is shown in Table \ref{tab:VLAobs}, and the location of the IFU with respect to the galaxy is shown in Figure~\ref{fig:rgbimage}. Due to time constraints and the high declination of the source, we obtain $3\times 1200\,\mathrm{s}$ `science' exposures of the source to maximize the signal-to-noise of the emission lines. Bias frames, Cu-Ar arcs, sky and dome flats, and exposures using a wire coronagraph were obtained at the beginning of the night to characterise the CCD and the alignment of the slitlets, with additional Cu-Ar arc and bias frames obtained between science exposures to account for wavelength drift and the changing temperature of the CCD over the course of the night. We also obtain one off-source frame to remove skyline contamination during the reduction process. Observations of a spectrophotometric standard star with a similar airmass are also obtained to perform photometric calibration and correct for telluric absorption.

Data reduction is performed using the PyWiFeS package \citep{PyWiFeS} to produce a spectrophotometrically calibrated and fully co-added data cube, and a corresponding error cube processed in parallel. The spectrum is then corrected for Milky Way dust reddening using \texttt{brutus}\footnote{http://fpavogt.github.io/brutus}. \texttt{brutus} applies a correction according to the \cite{Fitzpatrick1999} reddening law with $R_v=3.1$. Galactic extinction values used in the correction are obtained from the \cite{Schlafly11} recalibration of the \cite{Schlegel98} infrared dust map. All subsequent analysis is then carried out in the rest-frame.

We find the average signal-to-noise ratio of the stellar continuum in the WiFeS spectra too low to generate reliable fits to the stellar population. We extract resolved maps of several prominent emission lines from the WiFeS data cubes (Fig.~\ref{fig:emissionmap}) and calculate emission line ratios to study the distribution and characteristics of star-forming environment in \host{}, including gas phase metallicity. 
To estimate the age and metallicity of the stellar population of \host{} we utilize archival spectra from the Sloan Digital Sky Survey (SDSS) Data Release 16 \citep{SDSSlegacy,SDSSdr16}. SDSS spectra integrate light from a circular fiber aperture with a diameter of $\sim 3"$, centered on the core of the galaxy. As we do not precisely know the location of \frb{}, the stellar population observed using the integrated host galaxy light serves as a useful proxy to obtain limits on the possible delay time distribution of the progenitor. SDSS spectra are pre-corrected for Galactic dust reddening using the same reddening laws used for the WiFeS spectra, and only require to be shifted to rest-frame wavelengths for analysis.

\section{Results}

\subsection{Radio}\label{sec:radio}
\begin{figure*}
	 		\includegraphics[width=2\columnwidth]{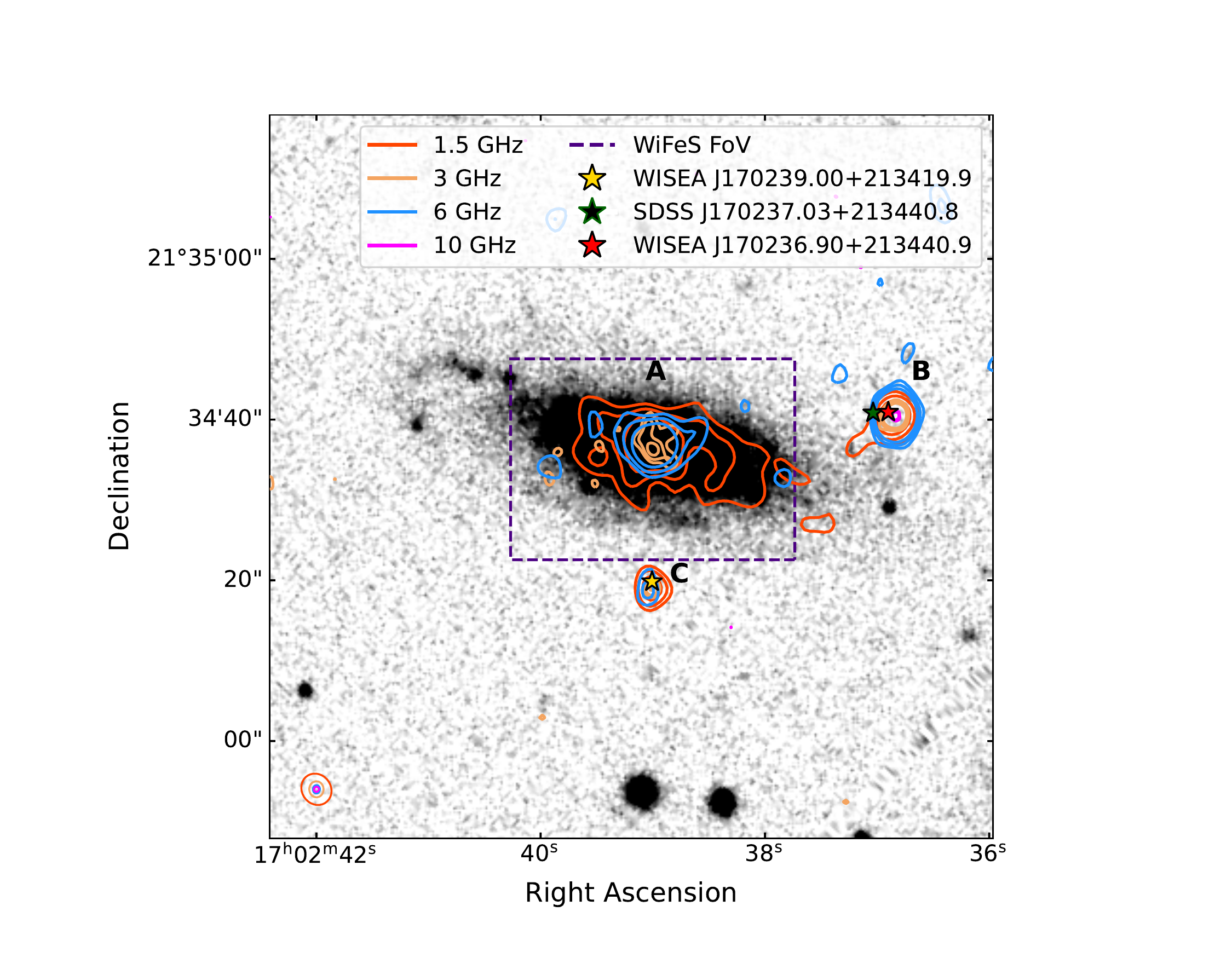}
	 		\caption{\label{fig:rgbimage} UGC10667 (background image: SDSS archive) with VLA contours overlaid. The central frequency of each VLA band is reported and the VLA beam size for each frequency band is shown in the bottom left corner. Concentric contour levels are $2\sigma\sqrt n$, where $n$ is the contour level and $\sigma$ is the RMS noise level of the image. We also show the three optical/IR sources with no cataloged redshift that are spatially coincident with VLA sources B and C, thought to be background AGN (see Section \ref{sec:radio}). The FoV of the WiFeS instrument is indicated by the dashed purple rectangle.}
\end{figure*}

The galaxy \host{} shows diffuse radio emission concentrated around a central radio source (see source `A' in Figures \ref{fig:rgbimage} and \ref{fig:VLA-Lband}). The radio emission from the central regions of the galaxy (see flux densities quoted in Table~\ref{tab:VLAobs}) gives a spectral index of $\alpha=-1.1\pm0.1$.

\begin{figure}
	 		\includegraphics[width=\columnwidth]{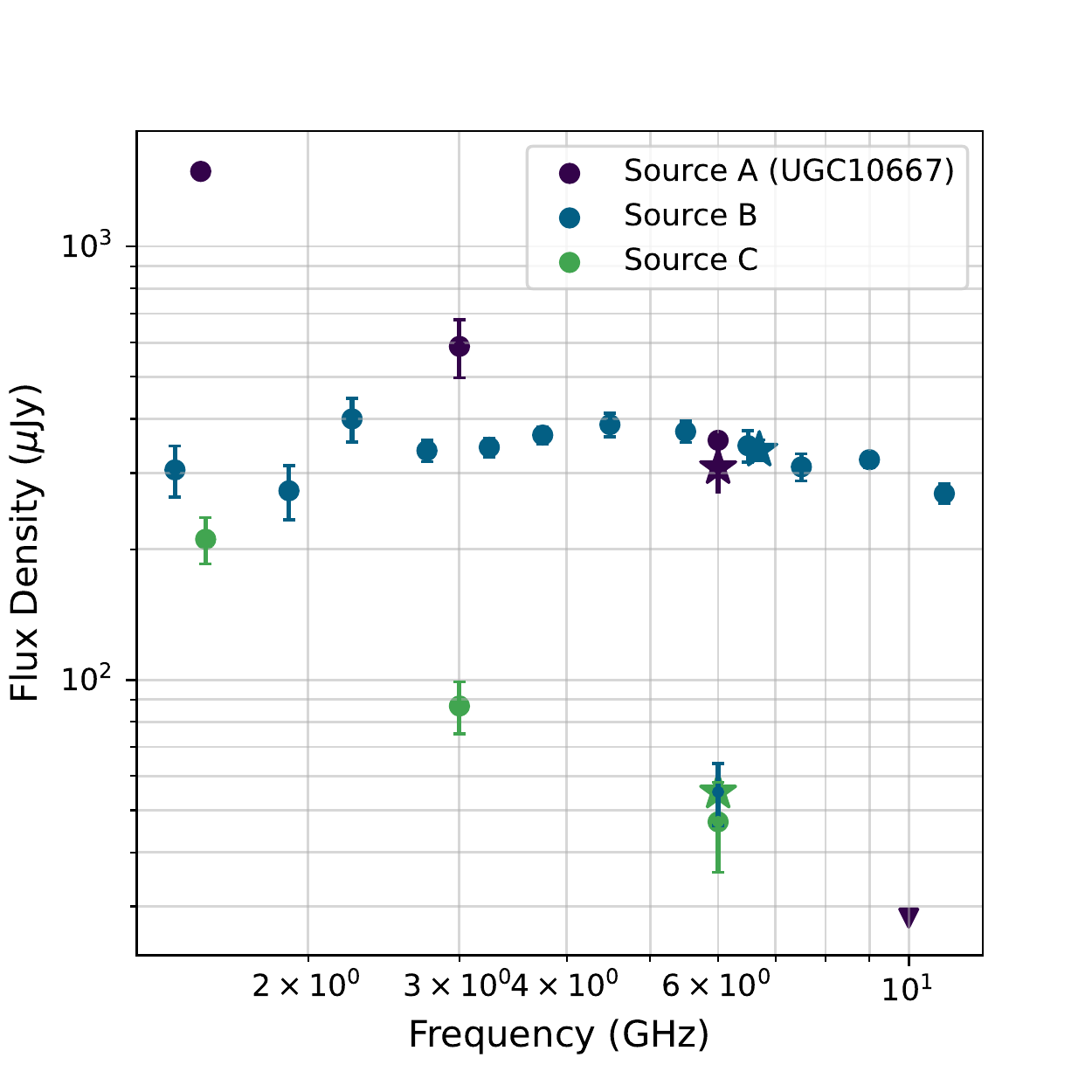} 		\caption{\label{fig:VLA-spectra}VLA radio spectra of sources labelled A, B, and C in Figure \ref{fig:VLA-Lband}. Circles show the spectra recorded on 2021-10-16 and stars show the 6\,GHz data recorded on 2021-09-09. 3$\sigma$ upper limits are indicated by inverted triangles.}\end{figure}

The 3 and 6\,GHz images from October 2021 show denser regions of radio emission in the disk of the galaxy, whilst the 1.5\,GHz image shows more diffuse radio emission throughout the disk. Radio continuum emission from galaxies generally arises at lower frequencies from non-thermal synchrotron emission associated with cosmic-ray electrons accelerated in the magnetic field of the galaxy, and at higher frequencies from thermal bremsstrahlung (free-free) emission around massive star-forming regions \citep[e.g.][]{Jong1985,Turner1994}. We deduce that much of the diffuse radio emission observed is due to active star formation within the galaxy. We infer the 1.4\,GHz star formation rate of the galaxy ($\rm{SFR}_{1.4\rm{GHz}}$) using \citet{Murphy2011}

\begin{align}
    \left(\frac{\rm{SFR}_{1.4\rm{GHz}}}{M_{\rm{\odot}}\,\rm{yr}^{-1}}\right) = 6.35 \times 10^{-29} \left(\frac{L_{\rm{1.4 GHz}}}{\rm{erg\,s^{-1}\,Hz^{-1}}}\right),
\end{align}

where $L_{\rm{1.4 GHz}}$ is the 1.4\,GHz luminosity of the galaxy. Based on the measured flux density of the galaxy at 1.5\,GHz, we infer a luminosity of $L_{\rm{1.4 GHz}}\approx3.67\times10^{28}$\,erg\,s$^{-1}$\,Hz$^{-1}$ (assuming a spectral index of $\alpha=-1.1$ for the galaxy, Figure \ref{fig:VLA-spectra}) and thus $\rm{SFR}_{1.4\rm{GHz}}\approx2.3\,M_{\rm{\odot}}\,\rm{yr}^{-1}$.
	 		
 In Figure \ref{fig:VLA-Lband}, there are two compact radio sources to the east (`B') and south (`C') of the nucleus of \host. The 1--10\,GHz radio spectra for each of these sources is plotted in Figure \ref{fig:VLA-spectra}. In the VLA 1--10\,GHz observations from 2021 October 16, the source labelled `B' (RA, Dec 17:02:36.8, +21.34.40.5) shows a relatively flat spectrum with flux density $\approx350$\,$\mu$Jy\,beam$^{-1}$; this flux density did not change between the September and October VLA 6\,GHz observations. The source spectrum shows a slight peak at approximately 5\,GHz.  In the VLA 1--10\,GHz observations from 2021 October 16, the source labelled `C' (RA, Dec 17:02:39.0, +21.34.18.95) shows a steep spectrum with $\alpha=-1.16$ and $F_{1.5\rm{GHz}}=211\pm26\,\mu$Jy\,beam$^{-1}$. There was no significant change in flux density of this source between the September and October VLA 6\,GHz observations.
 
We searched for optical counterparts that may be associated with VLA sources B and C using NED. A search for of all extragalactic sources in a 1 arcmin radius around \host\, returns 38 objects classified as galaxies or infrared sources (IrS). Of these, SDSS J170237.03+213440.8 ($m_r=22.856$) and IrS WISEA J170236.90+213440.9 are spatially coincident with B, and WISEA J170239.00+213419.9 ($m_r=22.308$) is spatially coincident with C. These sources are plotted alongside the VLA contours in Fig.~\ref{fig:rgbimage}. Given the catalog object density of one object per 0.083 square arcmin, these are likely true associations. SDSS J170237.03+213440.8 and IrS WISEA J170236.90+213440.9 are likely the same background galaxy, with separation of 0.94'' from eachother, and 2.7'' and 2.3'' from B respectively, while WISEA J170239.00+213419.9 is 1.8'' from C. There is no significant radio emission from any of the remaining 35 sources.
Given the peaked radio spectrum of source B (Figure \ref{fig:VLA-spectra}) and possible association with a faint optical source, we deduce this is likely a compact AGN. The WISE W1-W2 and W2-W3 colors and radio emission are consistent with a class of Infrared Faint Radio Sources identified in \cite{Patil2022}. Likewise, given the steep radio spectrum of source C (Figure \ref{fig:VLA-spectra}) and association with an optical counterpart, we conclude this is likely an old background AGN. In both cases, we find that the characteristics of the sources indicate that they are likely not associated with \host.

\subsubsection{Transient Search}\label{sec:tran_search}
We conducted a search for transient emission at 6\,GHz within 100\,arcseconds of the center of \host{} to look for an afterglow associated with \bns{} that has been putatively associated with \frb{}. We used the source detection software \texttt{Aegean} \citep{Hancock2012,Hancock2018} to search for sources in each of the September and October VLA 6\,GHz images. 
The transient search region is based on the known short gamma-ray burst (GRB) host offset range of $0.5-75$\,kpc \citep{fong13}, which we used as a proxy for where we might expect the BNS merger to be located. At a redshift of $z=0.031$, this corresponds to a maximum host offset of $\sim100$\,arcseconds from the centre of UGC10667. 

In the September image we identified four sources within the search radius, including \host{}, and in the October image we identified three sources within the search radius, including \host. There was a 4$\sigma$ detection of a source in the September image but not October with coordinates RA, Dec: 255.64545, 21.553322 and a flux density of 56$\pm13\,\mu$Jy\,beam$^{-1}$ (compared to 3$\sigma$ upper limit of $<30\,\mu$Jy\,beam$^{-1}$ in October). Interestingly, this same source was marginally detected at 3\,GHz in October with a flux density of 42$\pm16\,\mu$Jy\,beam$^{-1}$. Due to its low significance, we cannot be confident that this source is real, and if it is, the spectral shape and variability make it unlikely to be a radio afterglow of the BNS merger at 2.5\,yr post-burst (see Section \ref{sec:afterglow}). 
The remaining three sources that were detected in both 6\,GHz September and October images showed no statistically significant flux variation between the two epochs, excepting the host itself, which appeared brighter in the September image. We attribute this difference to be due to the change in VLA configuration from C in September to B in October causing a change in the amount of resolved emission that the interferometer was sensitive to. 

We thus conclude that no convincing transient emission was detected above a $5\sigma$ threshold of $>56\,\mu$Jy\,beam$^{-1}$ between 2021 September 9 and 2021 October 16 within 100\,arcsecond of \host{} at 6\,GHz. We note that this does not rule out transient radio emission in the 2.5 years preceding these observations so in Section~\ref{sec:afterglow} we discuss how our radio constraints compare to synthetic short GRB radio afterglow spectra and light curves.

\subsection{Optical}
\subsubsection{Full spectrum fitting and stellar population properties}
\begin{figure*}
	 		\includegraphics[width=\textwidth]{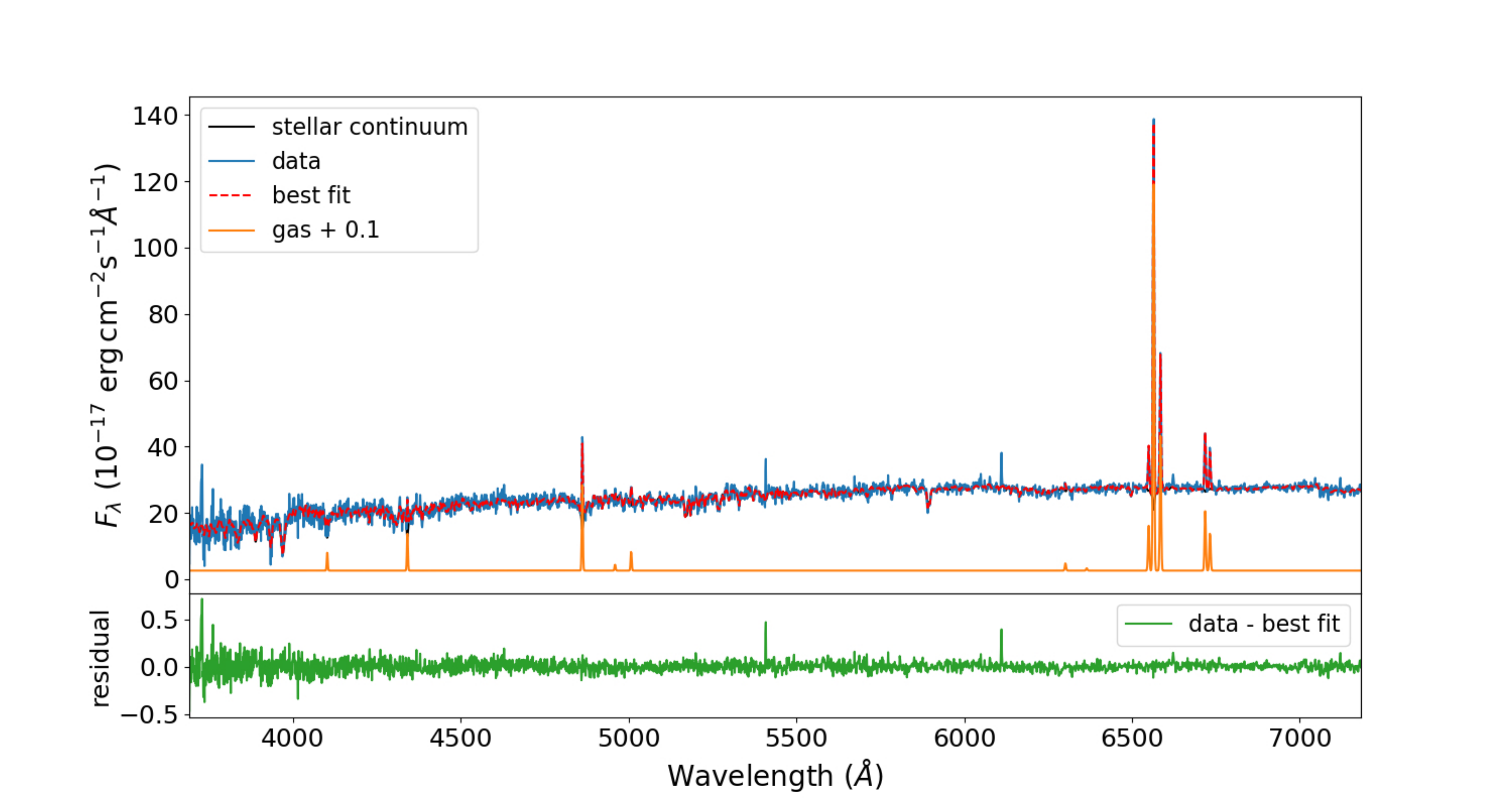}
	 		\caption{\label{fig:ppxfspec} Spectrum of UGC10667 observed by SDSS. \texttt{pPXF} fits SSP models to optical spectra, producing a best fit model that decomposes light into both nebular emission (orange, shifted by a constant factor for visibility) and stellar continuum (black). The best fitting model (red) is a composite of multiple MILES SSPs weighted according to Fig. \ref{fig:ppxfweights}. The best fitting gas model (orange) is used for the derivation of emission line ratios to remove contamination from stellar absorption in the H$_\beta$ line.}
\end{figure*}

\begin{figure}
	 		\includegraphics[width=\columnwidth]{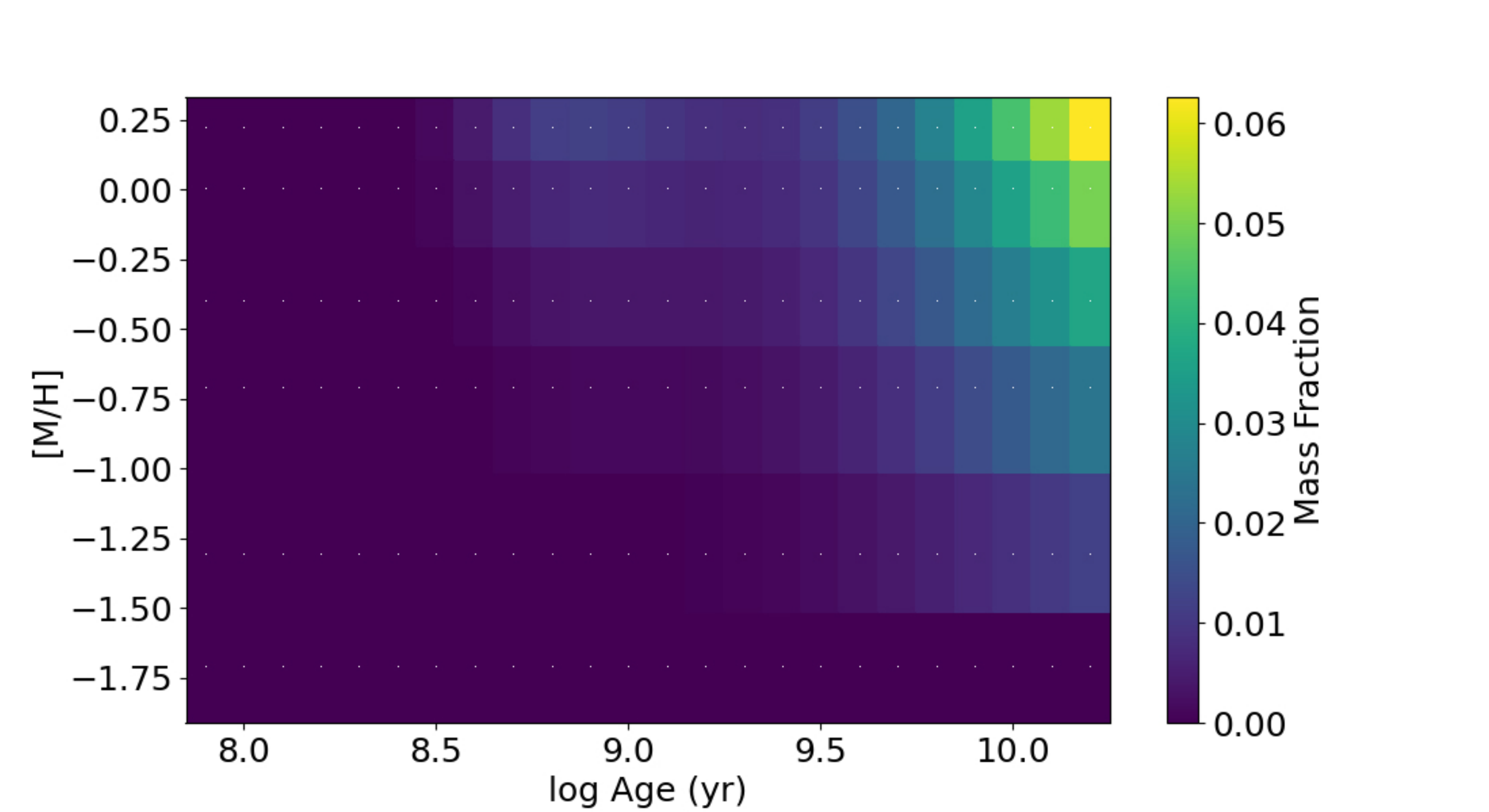}
	 		\caption{\label{fig:ppxfweights}Weights associated with each SSP template included in the \texttt{pPXF} analysis of the spectrum of UGC10667. The majority of the stellar mass is concentrated in the old stellar population,with an additional component with ages of $\sim 1\,\mathrm{Gyr}$. The color scale shows the relative mass of stars in each log-age - log-metallicity bin.}
\end{figure}

To estimate the age and metallicity of the stellar population of \host{} we utilize archival spectra from the Sloan Digital Sky Survey (SDSS) \citep{SDSSlegacy,SDSSDR6}, as the signal-to-noise ratio of the stellar continuum in the observations obtained using WiFeS is insufficient to carry out the full-spectrum fitting procedure. To understand the age of a stellar population, one must account for the degeneracy between the effects of stellar population age and metallicity on the observed spectrum. To do this, we use the Penalized Pixel Fitting (\texttt{pPXF}) method \citep{ppxf1, ppxf2} to perform simultaneous fits to a grid of simple stellar population models (SSPs). SSPs are drawn from the MILES library \citep{MILES}, which provides $985$ stellar spectra across a wavelength range of $3525-7500$\,\AA\,at 2.5\,\AA\,spectral resolution, for stellar metallicities ranging from $[M/H] = 0.22$ to $[M/H] = -2.3$. The stellar continuum and gas-phase emission lines are fit simultaneously in \texttt{pPXF}, with independent velocity components for the stars and gas. Intrinsic extinction that arises due to dust in the galaxy and calibration errors are included using a fifth-order multiplicative polynomial to approximate their impact on the stellar continuum. We apply regularization to ensure the best-fitting template weights are biased towards the smoothest solution still consistent with the data, choosing the regularization parameter iteratively. This is physically motivated in the context of how stellar populations evolve, and moreover motivated by the data as the problem is inherently ill-posed. Correct application of the regularisation parameter in \texttt{pPXF} enables the decomposition of bulge and disk stellar populations.\\ 
The resulting weights output by \texttt{pPXF} (Figure \ref{fig:ppxfweights}) that apply to each of the log-age - log-metallicity bins thus corresponds to the mass fraction of stars that contribute to the total flux observed. The results of the full spectrum fit, with each component of the fit is shown in \ref{fig:ppxfspec}. There are two components to the stellar population of \host{}: an older `bulge-like' population dominates the stellar mass with an age $> 10\,\mathrm{Gyr}$ and a younger stellar population with stars with ages $\sim 1\,\mathrm{Gyr}$ is also present.

\subsubsection{Gas phase line emission}
\begin{figure*}
	 		\includegraphics[width=\textwidth]{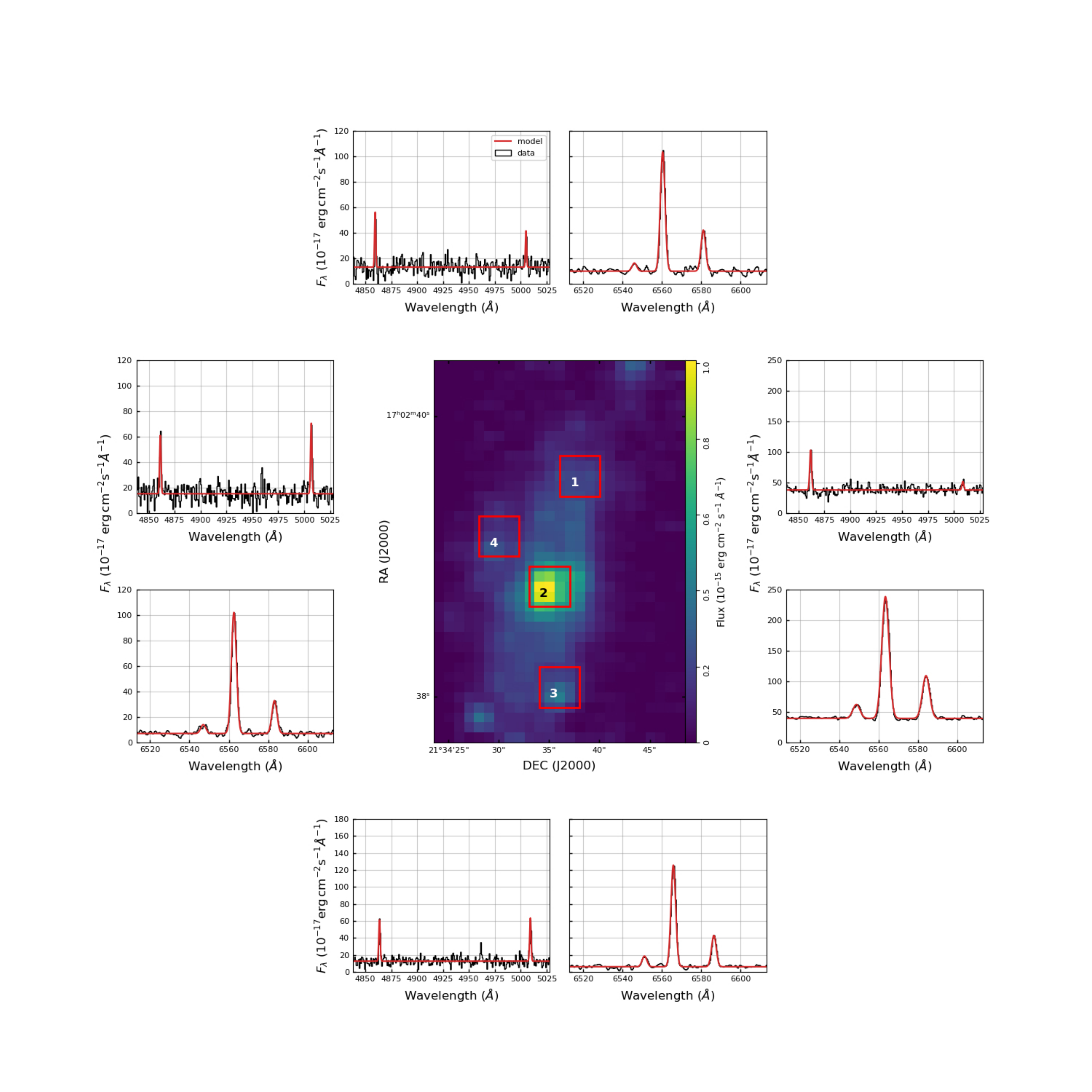}
	 		\caption{\label{fig:emissionmap} H$\alpha$ emission from WiFeS data cube (center). Each inset shows the strong emission lines observed in different regions of the galaxy: region 1 (top), region 2 (right), region 3 (bottom) and region 4 (left). Emission lines are fitted using a single-component Gaussian for each line and a constant term to account for the stellar continuum.}
\end{figure*}
There are strong emission lines present in the spectrum of \host{} (Fig. \ref{fig:ppxfspec} and Fig. \ref{fig:emissionmap}). We wish to understand the star formation rate of the host, characterise the regions of active star formation or other emission, and deduce the gas-phase metallicity to build a complete picture of \host\ as the putative host galaxy of \frb{}. 

To study the processes that give rise to the emission lines present in the spectrum of UGC10667, we select four regions of interest (ROI) in the galaxy (Fig.~\ref{fig:emissionmap}) representing regions of the disc (1, 3 and 4) and nucleus (2). Integrated spectra are extracted from each of these apertures (central panel of Fig.~\ref{fig:emissionmap}, with the inset spectra showing the strong lines in the red and blue regions of the optical spectrum for each region). We also include the SDSS spectrum used for the full-spectrum fitting procedure in the analysis, which integrates light from a circular aperture with a diameter of $3\,\mathrm{arcsec}$. For the SDSS spectrum, we use the gas spectrum derived by \texttt{pPXF} (Fig.~\ref{fig:ppxfspec}) to compute line fluxes and ratios to account for the absorption in the Balmer lines arising from features in the resolved stellar spectra. For the WiFeS spectra, we ignore this effect as the stellar continuum is undetected due to the noise in the observation.

In each ROI, we fit the strong lines of H$_\alpha$, H$_\beta$ \nii]$\lambda6584$ and \oiii]$\lambda5008$ and construct emission line ratios of $\log_{10}($\nii$]/H_\alpha)$ and $\log_{10}($\oiii$]/H_\beta)$. These ratios, listed in Table~\ref{tab:lines}, are used to construct a Baldwin, Phillips \& Terlevich (BPT) diagnostic diagram (Fig. \ref{fig:FRB_host_comp}) to investigate the processes that give rise to the strong emission lines. 
\begin{table*}
	\centering
	\caption{Emission line ratios associated with UGC10667 for each ROI defined in Fig \ref{fig:emissionmap}, and the SDSS spectrum used for the full-spectrum fitting procedure. Quoted errors are based on the observed variance of the line fluxes. Values where errors are not quoted are several order of magnitude smaller than the value - in particular for the high SNR SDSS spectrum. Errors in SFR and $12+\log(\mathrm{O/H})$ do not take into account systematic uncertainties associated with the methods used.}
	\label{tab:lines}
	\begin{tabular}{lccccc}
		\hline
		ROI & H$_\alpha$ flux ($10^{-15}\,\mathrm{erg\,cm^{-2}\,s^{-1}}$) & $\log_{10}($\nii]/H$_\alpha$)& $\log_{10}($\oiii]/H$_\beta$)	& SFR ($\mathrm{M_\odot\,yr^{-1}}$) & 12 + log(O/H)\\
		\hline
		SDSS & $5.93$ & $-0.48$ & $-0.6350$ & $0.039\pm0.01$ & 9.03 \\
		1 & $1.64\pm0.02$ & $-0.462\pm0.005$ & $-0.18\pm0.05$ & $0.018\pm0.005$ & $8.90\pm0.2$\\
		2  & $7.76\pm0.03$ & $-0.455\pm0.002$ & $-0.67\pm0.05$ & $0.086\pm0.02$ & $9.05\pm0.01$\\
		3 & $1.64\pm0.02$& $-0.511\pm0.004$ & $0.02\pm0.02$ & $0.018\pm0.005$ & $8.8\pm0.02$\\
		4 & $2.91\pm0.03$ & $-0.566\pm0.006$ & $0.08\pm0.02$ & $0.032\pm0.009$ & $8.77\pm0.04$ \\
		\hline
	\end{tabular}
\end{table*}

We derive the star formation rate in each of the ROI, and that derived from the SDSS spectrum, using the H$_\alpha$ line flux using a similar procedure to \cite{Heintz2020}. The SFR in $\mathrm{M_\odot\,yr^{-1}}$ is related to the intrinsic H$_\alpha$ line luminosity by 
\begin{equation}
\mathrm{SFR} = 4.98 \times 10^{-42} L_{\mathrm{H}_\alpha},
\end{equation}
where $L_{\mathrm{H}_\alpha}$ is given in $\mathrm{erg\,s^{-1}}$ and  $L_{\mathrm{H}_\alpha}= F_{\mathrm{H}_\alpha} \times 4\pi d_L$. We use the luminosity distance $d_L = 138.03 \pm 9.66 \,\mathrm{Mpc}$ listed by NED, using standard cosmology. This relation follows \cite{Kennicutt1998}. Consistent with \cite{Heintz2020}, we utilize the Chabrier IMF \citep{Chabrier2003} in this relation, assuming case B emission of \cite{OsterbrockandFerland} (i.e. the medium is optically thick to ionizing photons). The values of $F_{\mathrm{H}_\alpha}$ and the derived star formation rate for each region of interest are listed in Table \ref{tab:lines}. We find star formation rates of $\mathrm{SFR}\sim0.02 - 0.08\mathrm{M_\odot\,yr^{-1}}$ for each ROI. 

Finally, we infer the gas phase metallicity of the ROI using the strong emission lines. We consider the oxygen abundances given by $12 + \log(\mathrm{O/H})$ as a proxy for the gas-phase metallicity. We use the O3N2 calibration method, which is consistent with more direct methods that either use measurements of the electron temperature, or photoionization models \citep{Maiolino2019}. Specifically, the parametric form of \cite{Hirschauer2018} is: 
\begin{equation}
12 + \log(\mathrm{O/H}) = 8.987 - 0.297R - 0.0592R^2 - 0.0090R^3,
\end{equation}
where $R = \log_{10}(($\oiii$]/$H$_\beta)/($\nii$]/$H$_\alpha))$. The oxygen abundances for each region of interest are included in Table~\ref{tab:lines}.

\section{Discussion}
\subsection{\frb{} as an exceptional event}
\begin{figure}
	 		\includegraphics[width=\columnwidth]{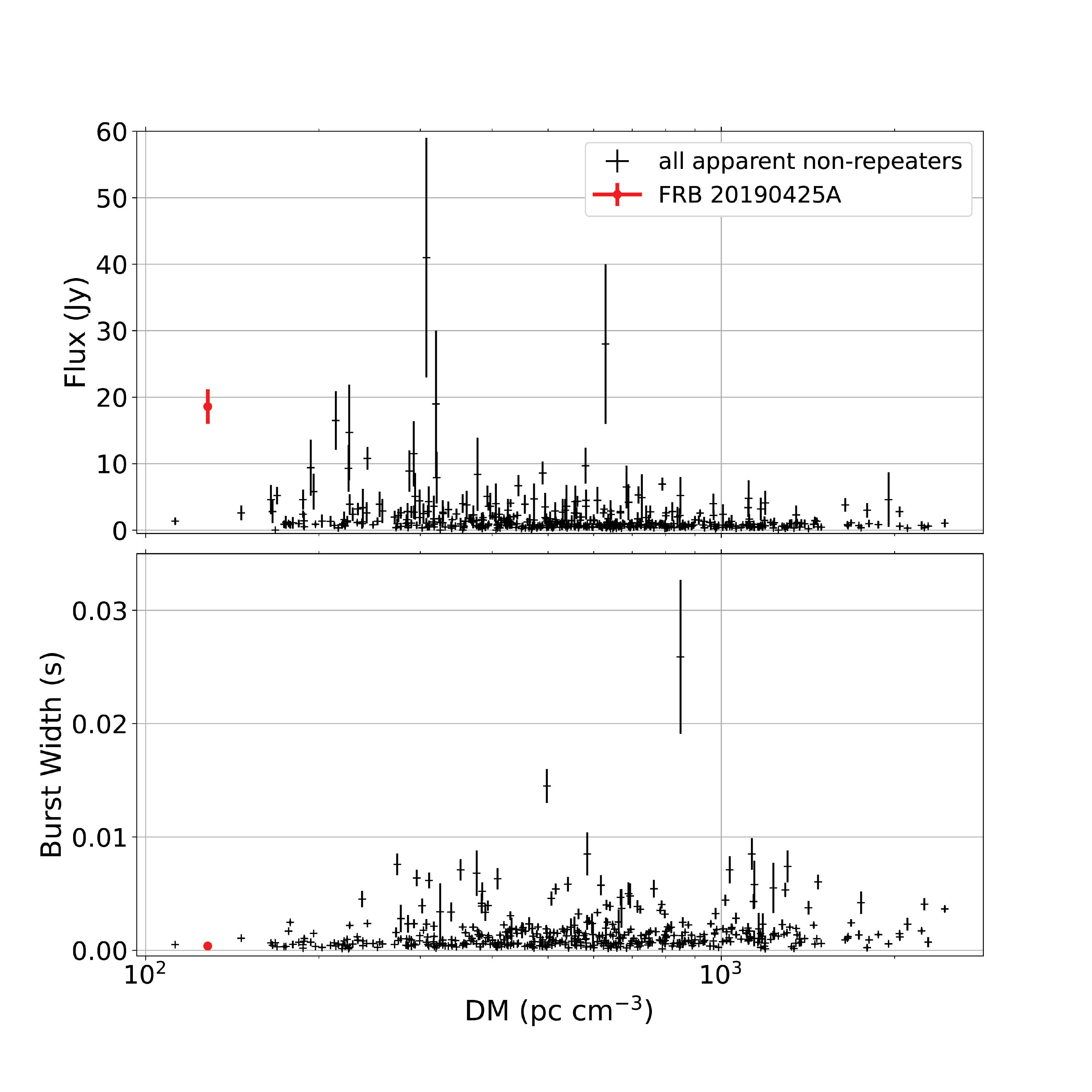}
	 		\caption{ \label{fig:Chimenonrepeates} Non-repeating FRBs identified in the CHIME/FRB Catalog 1. \frb\ (red marker) is noted for its short duration, high fluence, low DM and broadband frequency emission.}
\end{figure}

\frb{} is of particular interest as a low DM, bright, short duration FRB in its own right.
\begin{itemize}
    \item \textit{DM}: \frb{} is one of the three lowest DM apparent non-repeater events identified by CHIME/FRB Catalog 1.
    \item \textit{Flux/Fluence}: \frb{} exhibits the highest flux of the sample below a DM of $\sim 200\,\mathrm{pc\,cm^{-3}}$ (Fig.~\ref{fig:Chimenonrepeates} upper plot).
    \item \textit{Duration}: \frb{} has a very short pulse duration, with a fitted width of $0.3799 \pm 0.00023 \mathrm{ms}$
    (Fig.~\ref{fig:Chimenonrepeates} lower plot).
    \item \textit{Spectrum}: \frb{} is noted to exhibit a flat, broadband frequency spectrum with fluxes almost constant between $400-800\,\mathrm{MHz}$.
\end{itemize}
Comparing the properties of \frb{} to the 5\% of the CHIME/FRB catalog closest in DM, we identify three other events with fluences greater than that of \frb{}: FRB20190303B, FRB20190423A and FRB20190403E. We find that the duration of these events is substantially longer ($3-5 \,\mathrm{ms}$) and exhibit different frequency domain characteristics to the broadband emission seen in \frb{}. The fluxes of these events are lower than that of \frb{}. The high flux, short duration and flat spectrum of \frb{} was noted by Morioanu+2022 as consistent with the progenitor being a binary neutron star merger.

\subsection{Late-time multi-wavelength properties of any associated transient afterglow}\label{sec:afterglow}

The proposed BNS progenitor to \frb{} suggests there should be a synchrotron afterglow from an associated short GRB. This afterglow would be caused by its relativistic jets  ploughing into the ISM according to the internal-external (Fireball) shock model \citep{piran99}. However, its brightness and evolutionary properties would be highly dependent on our viewing angle with respect to the GRB jet axis,
which may or may not be choked and/or structured \citep{ghirlanda19}. 
In such cases, we need to wait until the jets slow and spread into our line-of-sight before afterglow emission becomes detectable. 

We first investigate whether the GW190425 BNS merger was viewed near/along the jet axis or whether the line of site had a higher inclination angle that was closer to the equatorial plane.
GW190425 was registered initially as a single-detector event in low-latency, detected by the LIGO Livingston observatory \citep{s190425z}, with a false alarm rate of 1 in $7 \cdot 10^4\,\mathrm{yr}$. Virgo observatory data was used for parameter estimation in the GWTC-2 analysis, however, the orientation of the BNS system cannot be constrained directly from the GW data \citep[$\cos(\iota) = 0.46^{+0.51}_{-1.42}$,][]{gwtc2}.

However, the possible association of \frb{} with \bns{} was interpreted by Morioanu et al.\ (2022) as the collapse of a supramassive, rapidly spinning neutron star into a black hole 2.5\,h post-merger that emitted an FRB likely isotropically \citep[see][for beaming angle considerations]{gourdji20} through the magnetic reconnection of the magnetosphere \citep{Zhang2014}. This
does suggest that this event was not observed close to the orbital plane of the merger, where merger ejecta might otherwise prevent the escape of the radio emission.
Additionally, simulations of GW190425 suggest a relatively low volume of NS material being ejected within a characteristic angle of $\sim10^{\circ}$--$20^{\circ}$ of the orbital plane \citep{Camilletti2022}, potentially allowing the emission of FRBs tens of degrees from the orbital axis to escape. Note that the small volume of NS material predicted by \citet{Camilletti2022} to be ejected is due to the NS equation of state chosen for use in this analysis predicting a rapid collapse to a black hole shortly after the merger, which would not be the case for the scenario of collapse after 2.5\,hr. However, simulations of GW170817, with a larger expected ejecta mass, still predict the bulk of the ejecta to be expelled in directions within $15^{\circ}$--$25^{\circ}$ of the orbital plane in most simulated scenarios \citep{Nedora2021}. 
As noted by these authors however, even the nature of the physical mechanism for expelling the material post-merger is extremely uncertain.
We therefore can only make the qualitative statement that an out-of-plane viewing angle is favoured, and that this does not necessarily imply an on-axis jet.


It is also not possible to completely rule out the prompt emission of gamma-rays by an on-axis jet from the BNS merger GW190425. Automated electromagnetic follow-up of the real-time detection of GW190425 \citep[S190425z,][]{GCN190425z} reported a marginally significant event \citep{INTEGRALGCN1} detected by the \textit{INTEGRAL} telescope's anti-coincidence shield system (SPI-ACS) \citep{SPIACS,Savchenko2012}. No significant signal was found by IBIS/PICsIT \citep{PICsIT}. A subsequent report details the event detected by SPI-ACS \citep{INTEGRALGCN2} as having a fluence $F = 2 \times 10^{-10}$ -- $2 \times 10^{-9}\,\mathrm{erg\,cm^{2}}$ in the $75-2000\,\mathrm{keV}$ range detected six seconds after GW190425. The significance of this event is less than $3\sigma$. The localization based on the SPI-ACS data is consistent with the location of \frb{}. The event detected by \textit{INTEGRAL} is likely to be consistent with noise than a real sGRB event associated with \bns{}.

While no gamma-ray emission was detected by the \textit{Fermi} Gamma-ray Burst Monitor (GBM) \citep{S190425zFermi} or the \textit{Swift} Burst Alert Telescope \citep[BAT,][]{SwiftGCN}, neither of these searches had coverage that overlapped with our proposed host \host{}. BAT only covered $\sim 7\%$ of the GW localization for \bns{} with $>10\%$ partial coding. It should be noted that the flux of the gamma-ray event detected by \textit{INTEGRAL} reported by \cite{INTEGRALGCN2} was below the three-sigma threshold for BAT events. In the case of the GBM non-detection, this instrument was only covering 55.6\% of \bns{} \citep{S190425zFermi, song19}, with the region of the sky consistent with the FRB event (much smaller than the GW localization area) located in GBM's Earth occultation region. This means that viewing angles derived from the GBM by \citet{song19} cannot be applied to our analysis.

Given the FRB association and the lack of  gamma-ray constraints on the jet orientation of \bns{} at the position of \host{}, we will assume the event was on-axis (i.e. our line-of-sight is close to or along any associated jet axis) for the purpose of this transient analysis where we search for synchrotron afterglow emission.
As mentioned in Section~\ref{sec:tran_search}, no convincing radio transients were found via a comparison between the two VLA 6 GHz epochs in September 2021 and October 2021 to a $5\sigma$ limit of $56\mu$Jy\,beam$^{-1}$.
In order to determine how constraining our VLA observations were for an on-axis jet at $2.5$\,y post-burst, we created synthetic spectra and light curves using the open-source {\sc Python} package \texttt{afterglowpy} \citep{ryan20}.  
These were computed assuming a 
short GRB median isotripic energy of $E_{K,{\rm iso}}=2 \cdot 10^{51}$\,erg \citep{fong15} and microphysical parameters of $\epsilon_e=0.1$, $\epsilon_B=0.01$, and $p=2.2$, which have been used for constraining the viewing angle and circumburst medium (CBM) densities of GW190814 from widefield radio counterpart searches \citep[e.g.][]{alexander21,dobie22}.  
Assuming a Top-Hat model for an on-axis jet with an opening angle of $10\deg$ \citep[][]{dobie19}, the synthetic spectra demonstrate that the 2021 September and October VLA 6\,GHz sensitivity of 
$56\mu$Jy\,beam$^{-1}$ ($5\sigma$)
was insufficient for detecting standard forward-shock radio afterglow emission for a typical short GRB CBM density of $n=10^{-3}$\,cm$^{-3}$. 
In fact, the CBM density would need to be $n>10^6$\,cm$^{-3}$ for the afterglow to be detectable at 6\,GHz at 2.5\,y post-burst, which is well beyond the range observed from known short GRBs \citep{fong15}. 
FRBs have been discovered to be embedded within compact, persistent radio sources, but even these environments have lower densities \citep[e.g. the persistent source in which the repeater FRB 121102 is embedded has a density of $n\approx10^2$\,cm$^{-3}$;][]{michilli18}. 
In addition, this late in the evolution of GW190425/FRB190425, we would not expect to detect much (if any) variability between the 2021 September and October epochs in the GHz frequency range. 
The synthetic light curves demonstrate that at $6$\,GHz (for $n=10^{-3}$\,cm$^{-3}$) only early-time observations would have detected the forward shock radio afterglow, which would have peaked at $\sim$2--3\,d post-burst at a flux density $>10^{3}\mu$Jy\,beam$^{-1}$, and remained detectable ($>56\mu$Jy\,beam$^{-1}$) for up to $\sim20$\,d post-burst.

No ultraviolet, optical or infra-red kilonova emission or an orphan afterglow (from an off-axis short GRB) associated with \bns{} was detected, despite fields containing \host{}, which had good coverage by a number of optical follow-up surveys \citep{Coughlin2019}. The limiting magnitude of the optical follow-up in the $3\,\mathrm{days}$ following the gravitational wave event is $m_{g, r}\sim 21$, and $m_K\sim 15$. Kilonovae are a particularly faint class of transient powered by the decay of r-process radionuclides \citep{Metzger10} --- at $z=0.03$ the peak magnitude of an event similar to GW170817 \citep{GW170817_multi} would be $m_r = 20 - 21$ in the r-band and fainter in g, making identification challenging. Moreover, the kilonova emission that may have been expected from an event like \bns{} is expected to be notably faint, much fainter than GW170817 \citep{Foley2020}. The kilonova emission is isotropic, with no significant dependence on viewing angle \citep{Kasen2017}, and the non-detection is not strongly constraining. An orphan afterglow may be substantially fainter again if the transient is being viewed off-axis \citep{Nakar2022}. While an on-axis afterglow would have been easily visible, we note that no confident detection of an short GRB associated with this event has been made. The rapid decline of the hypothetical kilonova or orphan afterglow lightcurve also makes long timescale followup difficult, as within $5-10\,\mathrm{days}$ the transient will fade well below detectable limits. The higher redshift of \bns{} in comparison to GW170817 can explain why no optical transient associated with this apparent coincidence is detected.
In summary, our late-time radio and optical analysis show no evidence of a transient associated with \frb{}/\bns{}.

\subsection{Comparison of UGC10667 to known FRB hosts}
\begin{figure*}
\begin{center}
\begin{tabular}{cc}
 \includegraphics[scale=0.4]{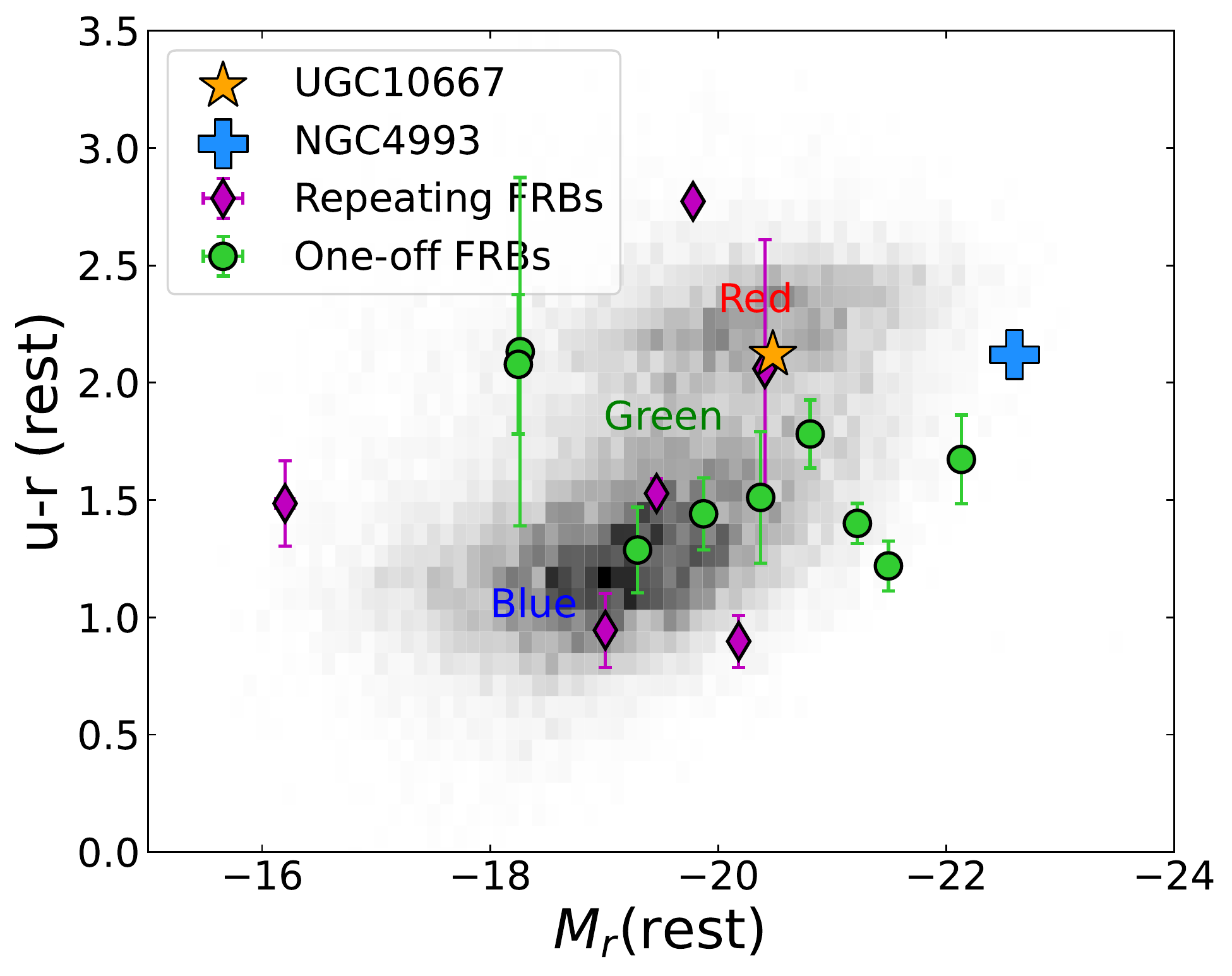} &
 \includegraphics[scale=0.4]{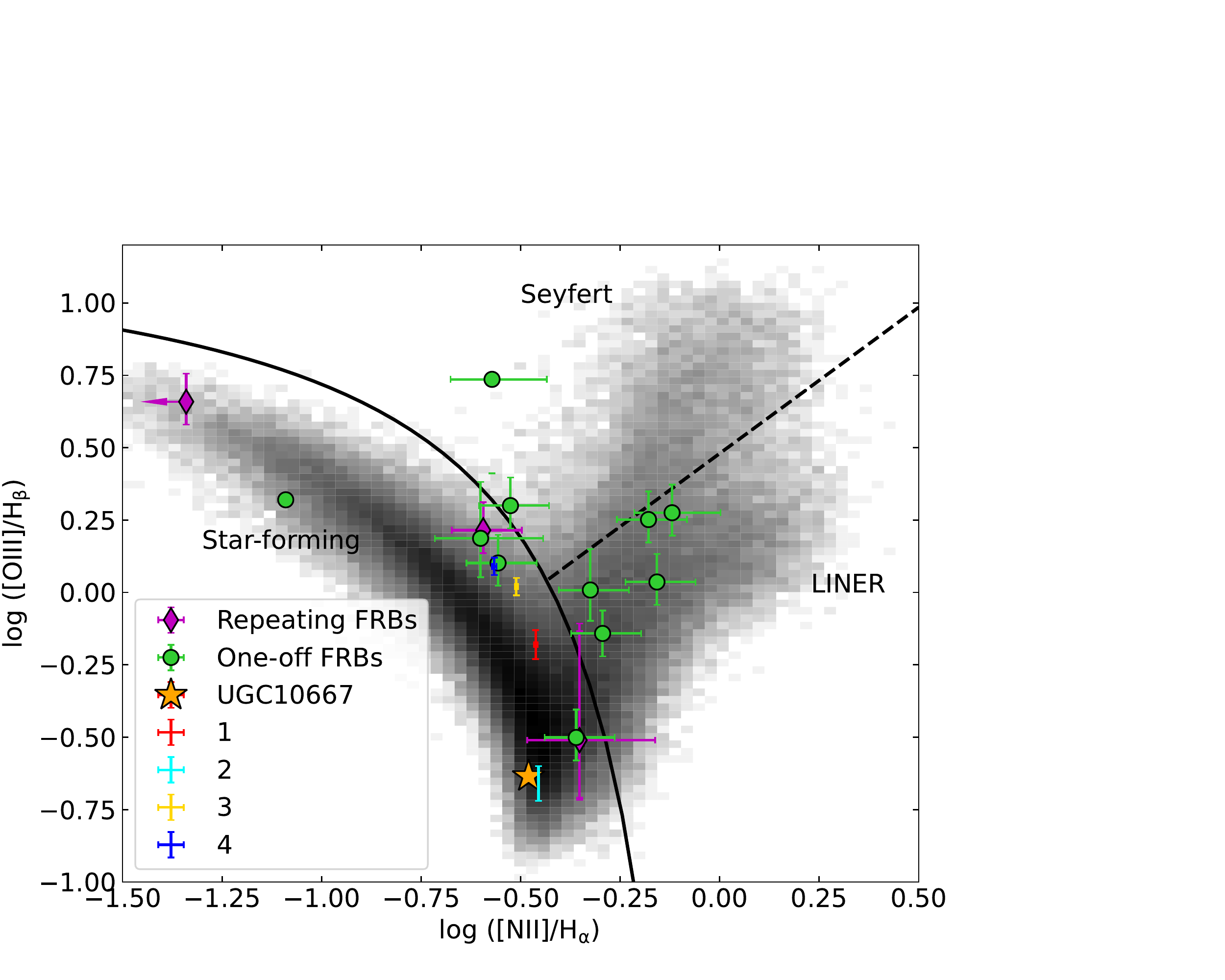}
 \end{tabular}
\caption{(Left): Rest-frame color-magnitude diagram comparing UGC10667 and other published FRB host galaxies \citep{Bhandari+22} to the PRIMUS galaxies population \citep{PRIMUS} at $z < 0.6$.
We note that the observed magnitudes for UGC10667 are taken from SDSS and are approximated as rest-frame magnitudes due to their low redshift and thus small k-correction. (Right): BPT classification  diagram for FRB hosts and UGC10667. 
The grayscale background depicts the density distribution of SDSS galaxies with redshifts ranging from $0.02$ to $0.4$. The dashed and dotted black lines demarcate the boundaries between SF galaxies and AGNs \citep{kauffmann03}, and between AGNs and LINERs, respectively \citep{fernandes10}. The region of interests within the UGC10667 galaxy are also over plotted. } 
\label{fig:FRB_host_comp}
\end{center}
\end{figure*}

\begin{figure}
\includegraphics[scale=0.4]{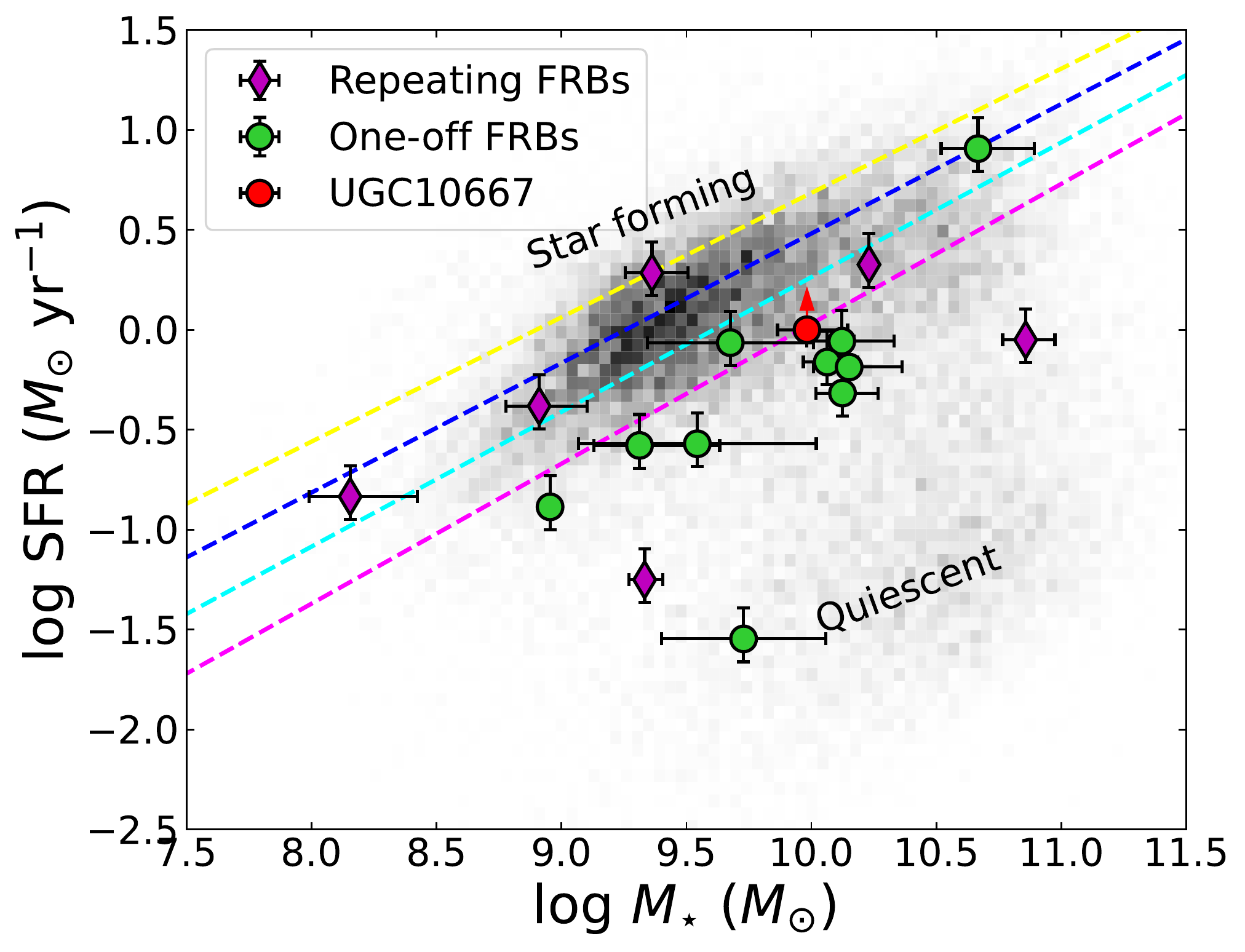} 	
\caption{Star-formation rate and stellar mass distributions of the host galaxies of repeating and non-repeating FRBs compared against a population of galaxies at $z<0.6$. Yellow, blue, cyan and magenta dashed lines represent the boundary separating the star-forming and quiescent galaxies and its evolution with redshift of $z=0,0.2,0.4, 0.6$ \citep{PRIMUS}. UGC10667 galaxy is shown as a red data point. }
\label{fig:SFR_mstar}
\end{figure}

As we do not know the precise localization of \frb, we can take the properties of the galaxy derived from the SDSS archival spectra as a good proxy to characterise the properties of the environment of \frb{} and compare them to other known FRB hosts. The observed emission line ratios, star formation rates and metallicities are found to be relatively consistent between the individual ROIs observed using the WiFeS IFU, with the most significant variation being in the \oiii]/H$_\beta$ emission line ratio (driven by variation in the flux of the \oiii] line, see Fig. \ref{fig:emissionmap}). In particular, we do not find any evidence for photoionization consistent with AGN or LINER activity in our optical spectra, including at the core of the galaxy where this may be expected, and the nebular gas emission is dominated by star formation. 

\host{} differs somewhat in its photometric properties when compared to the hosts of other one-off FRBs, although not exceptionally --- the majority of one-off FRB hosts are green valley galaxies, while \host{} is redder in color (Fig. \ref{fig:FRB_host_comp}), suggesting a large fraction of old stars (ages $>\mathrm{a\,few\,Gyr}$, consistent with the stellar population derived from the SDSS spectrum, in Fig.\ \ref{fig:ppxfweights}). It has remarkably similar photometric properties to the host of the repeating FRB~20201124a, although the host of FRB~20201124a has a slightly higher \nii]/\ha~ratio ($0.35$ compared to $0.45 - 0.5$ for \host{}).

All of the repeating FRB hosts sit on the star-forming side of the \cite{kauffmann03} diagnostic on the BPT diagram, but all except the host of FRB~20201124a are significantly bluer in $u-r$ color. In contrast, a large fraction of one-off FRB hosts exhibit higher \oiii]/H$_\beta$ ratios than observed in \host{}, indicative of LINER or AGN-like emission on the basis of the metrics of \cite{kauffmann03}. This may be suggestive of galaxies that contain a large amount of gas available for star formation or active accretion onto the central supermassive black hole (both of which can result in LINER-like emission). 

The oxygen abundances derived from the O3N2 diagnostics is consistent with that known of other FRB hosts. High metallicity rules out a common origin between \frb{} and classes of events such as long gamma-ray bursts that are known to preference sub-solar metallicity hosts \citep{Perley2016}. There is some evidence for a metallicity gradient in \host, with the nuclear spectra exhibiting higher values of $12+\log(\mathrm{O/H})$, however, the metallicities of even the disk ROIs are still super-solar.

The total SFR of \host{} derived from $1.4\,\mathrm{GHz}$ radio emission $2.3\,M_{\rm{\odot}}$, and Our WiFeS observations suggest the total SFR of \host{} is $\sim 1\,\mathrm{M_\odot\,yr^{-1}}$ in the full field-of-view. This indicates a SFR that is consistent with known FRB hosts \citep[$0.08-10\,\mathrm{M_\odot\,yr^{-1}}$,][]{Heintz2020,Bhandari+22}.

From archival SDSS spectra, we find no evidence of a substantial population of stars with ages $<1\,\mathrm{Gyr}$, with a SFR of $\sim 0.06\,\mathrm{M_\odot\,yr^{-1}}$ in innermost 3" of the galaxy. 

The disparity betweeen the WiFes full FoV-derived and SDSS-derived SFRs is driven by the differing FoV of the two instruments. WiFeS samples almost the entire host galaxy, whereas the SDSS fiber only samples a region 3" in diameter in the innermost regions of the galaxy. The four ROIs extracted from the WiFeS FoV suggest that the SFR is uniform across the galaxy. Nevertheless, the WiFeS full-FoV derived SFR values should be taken as a lower limit as the aperture may not capture the full extent of the star-forming gas as evidenced by radio emission (Fig.\ \ref{fig:emissionmap}). 

To determine the stellar mass of \host{}, fit SDSS photometry using the spectral energy distribution fitting program \texttt{CIGALE} \citep{Cigale}, and find $M_* \simeq 9.8 \times 10^{9}\,\mathrm{M_\odot}$. \host{} lies on the star-forming main sequence of galaxies, with a similar stellar mass and slightly higher star formation rate than other known hosts of one-off FRBs (Fig. \ref{fig:SFR_mstar}.)

Due to observational challenges including the high declination of the galaxy and the limited integration time of the observation, there is not sufficient signal-to-noise in the stellar continuum observed by WiFeS to perform full-spectrum fitting of the entire galaxy, thus there may be a bias toward finding an older stellar population in the central bulge regions ofthe galaxy that are sampled by SDSS. Nevertheless, the location of the galaxy on the color-color diagram (Fig. \ref{fig:FRB_host_comp}), last significant period of star formation being $\sim1$\,Gyr, and the lack of absorption features present in the optical spectra suggests a moderately long delay-time between SFR and the FRB progenitor event, such as would be expected in a BNS merger scenario \citep{Zhang2014}. However, clearly star forming activity has not completely finished in \host, and we cannot rule out that the progenitor of \frb\ arose from e.g.\ a young magnetar from a CCSNe \citep{Michilli2018_121102, bochenek2020}.

There is no clear observational evidence in our WiFeS IFU data that \host{} has undergone any recent major merger activity (for example, a significant population of young stars, LINER or AGN emission) that has been found in a number of other FRB host galaxies \citep[Ryder et al., in prep; ][]{Kaur2022_HI}. However, we do note that SDSS g-band images of \host{} exhibit a low surface brightness feature to the West of the WiFeS FoV (Fig \ref{fig:rgbimage}). Further observations of this feature are required to determine if it is kinematically distinct from \host{}, and is the result of a recent tidal interaction between the galaxy and a companion. Spectroscopic observations of this feature may also determine whether any significant population of young stars are present, further suggesting merger activity. We do note that this feature has no significant associated radio emission at $1.4\,\mathrm{GHz}$, hence any associated merger would likely be a dry merger, devoid of gas. 

While our spatially averaged properties are reasonable proxies for the properties of the galaxy as a whole -- and hence the likely local environment of the FRB itself -- we cannot rule out that the FRB occurred in a small and locally unusual region of the host, as was the case for FRB 20200120E, which was localised to a globular cluster within M81 \citep{Bhardwaj2021_M81,Kirsten2022}. Future observations to obtain deep, high-resolution IFU spectroscopy of \host{} to resolve the stellar continuum can enable the construction of a detailed star formation history of the galaxy. While the exact position of the FRB is unknown, the current work only utilizes SDSS spectra to obtain fits to the stellar continuum to extract stellar population age. As the SDSS fiber samples only the central regions of the galaxy, we do not necessarily resolve the full details of the stellar population and star formation history of the galaxy, which may shed further light on the nature of the FRB progenitor, or the recent merger history of this galaxy.

\subsection{Comparison to known sGRB/BNS merger host NGC4993}

It was suggested in \cite{Moroianu2022} that events such as \frb{} may be associated with a sub-population of FRBs that arise from compact object mergers --- hence the differing properties of \host{} in comparison to other one-off FRB hosts may support this theory.

\bns{} is the second BNS merger detected through the observation of gravitational waves. The first BNS merger, GW170817, was localized rapidly thanks to the serendipidous detection of sGRB emission by Fermi and INTEGRAL. The subsequent detection of kilonova emission on the outskirts of NGC4993 confirmed the massive elliptical galaxy as the host. If \host{} is indeed the host galaxy of \bns{}, it is pertinent to consider its properties in comparison to NGC4994.

The stellar population of NGC4993 is dominated by old stars \citep{Im2017}, and it exhibits photometric properties similar to \host{}. However, there is no evidence of ongoing active star formation, with the total SFR of massive stars in NGC4993 estimated to be $\sim 0.03\,\mathrm{M_\odot\,yr^{-1}}$ \citep{Sadler2017GCN}. The region local to the BNS merger GW170817 exhibits no evidence of recent star formation or a significant population of young stars \citep{Levan2017}. In contrast \host{} has a higher total SFR consistent with occupying the star-forming main sequence. Emission line ratios show that photoionization is driven by a low-ionization AGN in NGC4993 \citep{Sadler2017GCN, Levan2017}, and is similar in metallicity to other sGRB hosts and \host{} \citep{Contini2018}.

Being a quiescent galaxy with a low SFR, it is thought that the massive star progenitors of GW170817 formed at least $1\,\mathrm{Gyr}$ ago, with a lower limit on the age of the system that produced GW170817 of $200 - 400 \,\mathrm{Myr}$ \citep{Ebrova2020} based on kinematic modelling of a galaxy merger event that likely quenched star formation in NGC4993. It is interesting to note that there is substantial evidence that NGC4993 is the product of a galaxy merger, similar to many FRB host galaxies \citep{Heintz2020}. 

The stellar population properties of NGC4993 and \host{} are similar despite differences in their physical morphologies and star forming properties: they are both bulge dominated galaxies with $\sim 20\%$ of the stars having ages of $\sim 1\,\mathrm{Gyr}$. 

While we cannot entirely rule out young progenitors of the event that produced \frb{}, it is not out of the question that both galaxies may have hosted BNS events with similar progenitor channels.

\section{Conclusions}
\cite{Moroianu2022} propose a high probability of astrophysical coincidence between the LIGO/Virgo binary neutron star merger \bns{}, and the CHIME/FRB event \frb{}. The connection between these events can be explained by a neutron star merger, followed by the collapse of thee remnant: a rapidly-rotating, hypermassive neutron star. 

In this work we conducted two investigations: 
\begin{itemize}
    \item identifying the most probable host galaxy of \frb{} -- \host{} -- and a comprehensive multi-wavelength study of its properties of, and
    \item a search for afterglow emission from the putative associated GW event \bns{}.
\end{itemize} 

We find that \host{} has an almost 80\% probability of being the host of \frb{} using the PATH methodology of \cite{PATH}, with the next most probable host having only a 3\% chance of association. \host{} is a relatively massive galaxy ($M_* \simeq 9.8 \times 10^{9}\,\mathrm{M_\odot}$, based on fitting SED models to the SDSS photometry) with a modest star formation rate of $\sim 1-2 \,\mathrm{M_\odot\,yr^{-1}}$. The nebular optical and radio emission from the galaxy arises from star formation only. We find no evidence for a significant population of young stars, instead noting from full-spectrum fitting of optical SDSS data and the red color of the galaxy that it is dominated by stars with ages $>1\,\mathrm{Gyr}$. Although the ongoing star formation means we cannot categorically rule out a young magnetar as the origin of \frb{}, the characteristics of \host{} are consistent with a long-delay time origin for the transient, including the merger of a binary neutron star. 

From archival data and our observations made in September/October 2021, we find no evidence for radio afterglow associated with the FRB or GW event within $100\,\mathrm{kpc}$ of \host{}, however this is not inconsistent with observations made 2.5 years post-prompt phase. If \bns{} is indeed associated with \frb{}, we constrain the circumburst density to be $<10^5\,\mathrm{cm^{-3}}$ based on our non-detection of any afterglow emission. 

Irrespective of the putative coincidence between \bns{} and \frb{}, our identification of \host{} as the most likely host of \frb{} demonstrates that localizing low-DM FRBs identified by CHIME/FRB and similar instruments with large localization uncertainties to individual hosts is possible: it is possible to add more detail to the growing landscape of FRB host galaxies, even in the absence of arcsecond localizations.

\section*{Acknowledgements}

We acknowledge the rightful owners of the land this research was conducted on -- the Whadjuk (Perth region) Noongar people -- and pay our respects to elders past, present and emerging.

FHP is supported by funding from Australian Research  Council  Centre  of  Excellence for  Gravitational  Wave  Discovery (OzGrav) under grant CE170100004. GEA is the recipient of an Australian Research
Council Discovery Early Career Researcher Award (project number
DE180100346). SB is supported by a Dutch Research Council (NWO) Veni Fellowship (VI.Veni.212.058). AJG is supported by the Australian government through the Australian Research Council’s Discovery Projects funding scheme (DP200102471). NHW is supported by an Australian Research Council Future Fellowship (project number FT190100231). The contribution of CWJ was partially supported by the Australian Government through the Australian Research Council's Discovery Projects funding scheme (project DP210102103). MK was supported by the University of Western Australia and funded by the Australian Research Council (ARC) Centre of Excellence for Gravitational Wave Discovery OzGrav under grant CE170100004    

FHP thanks Henry Zovaro, Bruce Gendre, Eric Howell and David Coward for their useful comments and discussion on the analysis of the optical data. GEA thanks Geoffrey Ryan for discussion on use of the AfterglowPy package. FHP and AJG thank Freya North-Hickey for her work on the VLA DDT proposal.

This research has made use of the NASA/IPAC Extragalactic Database (NED),
which is operated by the Jet Propulsion Laboratory, California Institute of Technology,
under contract with the National Aeronautics and Space Administration.

The National Radio Astronomy Observatory is a facility of the National Science Foundation operated under cooperative agreement by Associated Universities, Inc.

This scientific work makes use of the Murchison Radio-astronomy Observatory, operated by CSIRO. We acknowledge the Wajarri Yamatji people as the traditional owners of the Observatory site. Support for the operation of the MWA is provided by the Australian Government (NCRIS), under a contract to Curtin University administered by Astronomy Australia Limited. We acknowledge the Pawsey Supercomputing Centre which is supported by the Western Australian and Australian Governments.

The Australian Square Kilometre Array Pathfinder is part of the Australia Telescope National Facility (https://ror.org/05qajvd42) which is managed by CSIRO. Operation of ASKAP is funded by the Australian Government with support from the National Collaborative Research Infrastructure Strategy. ASKAP uses the resources of the Pawsey Supercomputing Centre. Establishment of ASKAP, the Murchison Radio-astronomy Observatory and the Pawsey Supercomputing Centre are initiatives of the Australian Government, with support from the Government of Western Australia and the Science and Industry Endowment Fund. We acknowledge the Wajarri Yamatji people as the traditional owners of the Observatory site.

This work uses spectroscopic and photometric data obtained by the Sloan Digital Sky Survey (SDSS), specifically from the SDSS-III release. Funding for SDSS-III has been provided by the Alfred P. Sloan Foundation, the Participating Institutions, the National Science Foundation, and the U.S. Department of Energy Office of Science. The SDSS-III web site is http://www.sdss3.org/.

SDSS-III is managed by the Astrophysical Research Consortium for the Participating Institutions of the SDSS-III Collaboration including the University of Arizona, the Brazilian Participation Group, Brookhaven National Laboratory, Carnegie Mellon University, University of Florida, the French Participation Group, the German Participation Group, Harvard University, the Instituto de Astrofisica de Canarias, the Michigan State/Notre Dame/JINA Participation Group, Johns Hopkins University, Lawrence Berkeley National Laboratory, Max Planck Institute for Astrophysics, Max Planck Institute for Extraterrestrial Physics, New Mexico State University, New York University, Ohio State University, Pennsylvania State University, University of Portsmouth, Princeton University, the Spanish Participation Group, University of Tokyo, University of Utah, Vanderbilt University, University of Virginia, University of Washington, and Yale University. 

The ANU 2.3m telescope is operated by the Australian National University. We acknowledge the traditional owners of the land on which the ANU 2.3m telescope stands, the Gamilaraay people, and pay our respects to elders past and present. FHP acknowledges Siding Spring Observatory director Chris Lidman for support to obtain optical observations of UGC10667 at his discretion.

\section*{Data Availability}
Data products associated with this work can be supplied by the authors on reasonable request.



\bibliographystyle{mnras}
\bibliography{FRB_bib} 


\bsp	
\label{lastpage}
\end{document}